\documentclass[fleqn,usenatbib]{mnras}

\usepackage{newtxtext,newtxmath}

\usepackage[T1]{fontenc}

\DeclareRobustCommand{\VAN}[3]{#2}
\let\VANthebibliography\thebibliography
\def\thebibliography{\DeclareRobustCommand{\VAN}[3]{##3}\VANthebibliography}

\usepackage{graphicx}	
\usepackage{amsmath}	

\usepackage{xspace}     

\usepackage{tabularx}   

\newcommand{\tess}{\textit{TESS}\xspace}
\newcommand{\cheops}{\textit{CHEOPS}\xspace}
\newcommand{\kepler}{\textit{Kepler}\xspace}
\newcommand{\corot}{\textit{CoRoT}\xspace}

\newcommand{\sbart}{\texttt{S-BART}\xspace}
\newcommand{\pyaneti}{\texttt{pyaneti}\xspace}

\newcommand{\logrhk}{$\log{R'_\mathrm{HK}}$\xspace}

\title[Gliese 12\,b]{The mass of the exo-Venus Gliese 12\,b, as revealed by HARPS-N, ESPRESSO, and CARMENES}

\author[D. A. Turner, Y. N. E. Eschen, F. Murgas et al.]{
Daisy A. Turner$^{1}$\thanks{E-mail: dat936@student.bham.ac.uk},
Yoshi Nike Emilia Eschen$^{2}$,
Felipe Murgas$^{3,4}$,
Annelies Mortier$^{1}$,
Thomas G Wilson$^{2}$,
\newauthor
Jorge Fern\'andez Fern\'andez$^{2}$,
Nicole Gromek$^{5}$,
Giuseppe Morello$^{6,7}$,
Hugo M. Tabernero$^{8,9}$,
Jo Ann Egger$^{10}$,
\newauthor
Shreyas Vissapragada$^{11}$,
Jos\'e A. Caballero$^{12}$,
Stefan Dreizler$^{13}$,
Alix Violet Freckelton$^{1}$,
Artie P. Hatzes$^{14}$,
\newauthor
Ben Scott Lakeland$^{1}$,
Evangelos Nagel$^{13}$,
Luca Naponiello$^{15}$,
Siegfried Vanaverbeke$^{16,17}$,
Alexander Venner$^{18}$,
\newauthor
Mar\'ia Rosa Zapatero Osorio$^{12}$,
Pedro J. Amado$^{6}$,
V\'ictor J. S. B\'ejar$^{3,4}$,
Aldo Stefano Bonomo$^{15}$,
\newauthor
Lars A. Buchhave$^{19}$,
Andrew Collier Cameron$^{20}$,
Ilaria Carleo$^{15}$,
Priyanka Chaturvedi$^{14,21}$,
Ryan Cloutier$^{5}$,
\newauthor
Mario Damasso$^{15}$,
Mangesh Daspute$^{22}$,
Shishir Dholakia$^{18}$,
Sjoerd Dufoer$^{16}$,
Xavier Dumusque$^{23}$,
\newauthor
Aldo Fabricio Martinez Fiorenzano$^{24}$,
Adriano Ghedina$^{24}$,
Avet Harutyunyan$^{24}$,
Enrique Herrero$^{9}$,
\newauthor
Ancy Anna John$^{1}$,
Jorge Lillo-Box$^{12}$,
Nicolas Lodieu$^{3,4}$,
Mercedes L\'opez-Morales$^{25}$,
Luca Malavolta$^{26,27}$,
\newauthor
Luigi Mancini$^{28,15,29}$,
Giacomo Mantovan$^{26,27}$,
David Montes$^{30}$,
Juan Carlos Morales$^{8,9}$,
Belinda Nicholson$^{18}$,
\newauthor
Jaume Orell-Miquel$^{31}$,
Larissa Palethorpe$^{32,33}$,
Enric Palle$^{3,4}$,
Andreas Quirrenbach$^{34}$,
Sabine Reffert$^{34}$,
\newauthor
Ansgar Reiners$^{13}$,
Ignasi Ribas$^{8,9}$,
Ken Rice$^{32,33}$,
André M. Silva$^{35,36}$,
Alessandro Sozzetti$^{15}$,
\newauthor
Manu Stalport$^{37,38}$,
Lev Tal-Or$^{22,39}$,
Trifon Trifonov$^{34,40}$,
St\'ephane Udry$^{23}$,
Mathias Zechmeister$^{13}$
\\
Affiliations are listed at the end of the paper
}

\date{Accepted XXX. Received YYY; in original form ZZZ}

\pubyear{\the\year{}}

\begin{document}
\label{firstpage}
\pagerange{\pageref{firstpage}--\pageref{lastpage}}
\maketitle

\begin{abstract}
Small temperate planets are prime targets for exoplanet studies due to their possible similarities with the rocky planets in the Solar System. M dwarfs are promising hosts since the planetary signals are within our current detection capabilities. Gliese 12\,b is a Venus-sized temperate planet orbiting a quiet M dwarf. We present here the first precise mass measurement of this small exoplanet. We performed a detailed analysis using HARPS-N, ESPRESSO, and CARMENES radial velocities, along with new and archival \tess, \cheops, and MuSCAT2/3 photometry data. From fitting the available data, we find that the planet has a radius of $R_\mathrm{p} = 0.93\pm0.06 \,\mathrm{R_\oplus}$ and a mass of $M_\mathrm{p} = 0.95^{+0.29}_{-0.30} \,\mathrm{M_\oplus}$ (a $3.2\sigma$ measurement of the semi-amplitude $K=0.67\pm0.21\,\mathrm{m\,s^{-1}}$), and is on an orbit with a period of $12.761418^{+0.000060}_{-0.000055}\,\mathrm{d}$. A variety of techniques were utilised to attenuate stellar activity signals. Gliese 12\,b has an equilibrium temperature of $T_\mathrm{eq}=317 \pm 8\,\mathrm{K}$, assuming an albedo of zero, and a density consistent with that of Earth and Venus ($\rho_\mathrm{p}=6.4\pm2.4\,\mathrm{g\,cm^{-3}}$). We find that Gliese 12\,b has a predominantly rocky interior and simulations indicate that it is unlikely to have retained any of its primordial gaseous envelope. The bulk properties of Gliese 12\,b place it in an extremely sparsely populated region of both mass--radius and density--$T_\mathrm{eq}$ parameter space, making it a prime target for follow-up observations, including Lyman-$\alpha$ studies.

\end{abstract}

\begin{keywords}
planets and satellites: detection  – planets and satellites: terrestrial planets – planets and satellites: interiors – techniques: radial velocities – stars: individual: Gliese 12 – planets and satellites: individual: Gliese 12\,b
\end{keywords}



\section{Introduction}

One of the principal motivators in the field of exoplanetary science is the desire to find planets that are analogous to Earth or other small Solar System planets, in size, mass, composition, and potential habitability. M dwarfs have emerged as especially promising targets, as their relatively low masses and radii facilitate the detection of small planets. Their habitable zones also lie closer in, meaning potentially habitable planets are on more detectable orbits. While there could be drawbacks with habitability for planets hosted by M dwarfs (e.g. lower UV flux, tidal locking), the possibility of life is not ruled out for these planets \citep[see][for a review]{Shields2016}.

Significant progress has been made in identifying small planets, primarily through large surveys. Since the first light of pioneering telescopes such as \corot \citep[COnvection, ROtation and planetary Transits;][]{Baglin2003} and \kepler \citep{Borucki2010}, the detection of exoplanets has undergone a revolution, resulting in a dramatic increase in the rate of discoveries. These missions have demonstrated that small planets (here loosely defined as any planet with a radius smaller than $2$\,R$_\oplus$) are abundant, particularly around M dwarfs \citep[e.g., ][]{Dressing2015, Sabotta2021}.

As of May 2025, around 6000 exoplanets had been detected,\footnote{\url{https://exoplanetarchive.ipac.caltech.edu}} with thousands of other targets identified as exoplanetary candidates, evidencing the vital part that space-based surveys play in establishing a selection of targets for follow-up.
However, while these missions have revealed many small planets, the understanding of this population is incomplete without measurements of their masses.

Obtaining precise planetary mass measurements remains challenging and time consuming. This is especially true for planets with radial velocity (RV) semi-amplitudes of less than $1\,\mathrm{m\,s^{-1}}$ \citep[e.g.,][]{Pepe2011, Wilson2022, Stalport2025}, a regime characteristic of Earth-mass planets around M dwarfs \citep[e.g.,][]{Demangeon2021, Bonomo2023, Murgas2023, Basant2025}. Such signals are further obscured by stellar activity, a considerable source of noise that is typically pronounced for faint M dwarfs such as Gliese 12. To achieve the precision necessary to characterise Earth-mass planets, data from high-resolution spectrographs, such as HARPS-N \citep[High Accuracy Radial velocity Planet Searcher for the Northern Hemisphere;][]{Cosentino2012}, ESPRESSO \citep[Echelle Specrograph for Rocky Exoplanets and Stable Spectroscopic Observations;][]{Pepe2014, Pepe2021}, and CARMENES \citep[Calar Alto high-Resolution search for M dwarfs with Exoearths with Near-infrared and optical Echelle Spectrographs;][]{Quirrenbach2014, Quirrenbach2018}, are required. 

Currently, very few Earth-sized planets have both precise mass and radius measurements. This small sample of planets acts as a critical benchmark to understand planet composition, formation, evolution, and habitability. With measurements of both radius and mass, the bulk density of a planet can be calculated, allowing for an understanding of its approximate interior structure and composition. We can furthermore use the mass and radius to simulate scenarios for the atmospheric evolution of a planet and infer its prospects for atmospheric characterisation with, e.g., \textit{JWST} or \textit{HST}. 
Accurate atmospheric parameters can only be obtained if the mass of the planet is known to a sufficient precision. With no measurement of a planet's mass, only tenuous approximations can be made, as the solutions for the envelope composition are degenerate \citep{Batalha2019}.
Therefore, with more planetary information available, a planet's similarity to Solar System planets can more easily be gauged.

The transit of Gliese 12\,b was discovered by \tess (TOI-6251), and using separate validation data, \citet{Dholakia2024} and \citet{Kuzuhara2024} announced the validation of this planet. Gliese 12\,b is a planet with a radius close to $1$\,R$_\oplus$, putting it firmly in the Earth-like regime. It orbits its host star every $\sim$12.76\,d. Since Gliese 12 is an M dwarf, this makes Gliese 12\,b a temperate planet just outside the habitable zone as defined by \citet{Kopparapu2014}.

While \citet{Dholakia2024} and \citet{Kuzuhara2024} did use RV measurements to validate the planet, its mass remained unmeasured due to an insufficient amount of data. This lack of data left key aspects unconstrained, such as its density, composition and potential for life. In this study, we present the first mass measurement of Gliese 12\,b using extremely precise RVs from three spectrographs: HARPS-N, ESPRESSO and CARMENES.

In Section \ref{sec:data}, we describe the data sets used in this work. This includes both photometry and spectroscopy. The host star and its properties are described in Section \ref{sec:star}. We performed an individual photometry fit, an RV fit informed by the photometry fit, and a joint fit, all of which are described in Sections \ref{sec:Photonly}, \ref{sec:RVonly}, and \ref{sec:joint}. Finally, we discuss the implications of this new mass measurement in Section \ref{sec:discussion} and conclude in Section \ref{sec:conclude}.

\section{Data}
\label{sec:data}

\subsection{Photometry}

Gliese 12 has been observed extensively, as described by \citet{Dholakia2024} and \citet{Kuzuhara2024}. In this work, we use transit data from \tess, \cheops, MuSCAT2, MuSCAT3, and long-term photometry from ASAS-SN, LCOGT, TJO, and E-EYE.

\subsubsection{\tess}

The Transiting Exoplanet Survey Satellite \citep[\tess;][]{Ricker2015, Ricker2021} has observed Gliese 12 in a total of five sectors, each being approximately 27\,d long. Four of these sectors were previously presented by \citet{Dholakia2024} and \citet{Kuzuhara2024}, totalling around 135\,d of coverage. The orbital period of Gliese 12\,b was initially ambiguous, as the first few sectors of \tess data were consistent with periods of 12.76 or 25.52\,d. Ultimately, in sectors 42, 43, 57, and 70, a total of five transits were identified, leading to the confirmation of a planetary orbital period of $\sim$12.76\,d.

Gliese 12 was observed again in \tess sector 84, the first sector of Cycle 7, between 2024 October 01 and 2024 October 26 (BJD 2460584.50 -- BJD 2460609.50). These data were released after the publication of \citet{Dholakia2024} and \citet{Kuzuhara2024}, and two additional transits events were captured. These data were taken using CCD 3 on camera 1. In all, the data span approximately 105\,d.
We used the PDCSAP data, downloaded from the MAST portal\footnote{\url{https://mast.stsci.edu/}}.

\subsubsection{\cheops}
The CHaracterising ExOPlanet Satellite \citep[\cheops;][]{Benz2021} is a follow-up mission launched in 2019 with the goal to derive radii of planets more precisely. \citet{Dholakia2024} presented five \cheops visits of the transits of Gliese 12\,b that were obtained in the \cheops AO-4 guest observers programmes with the IDs ID:07 (PI: Palethorpe) and ID:12 (PI: Venner). Since then an additional visit has been obtained within the ID:12 programme and a transit observed. The six available \cheops visits are summarised in \autoref{tab:cheops}.

As Gliese 12 is a rather faint target to be observed by \cheops, we re-extracted the photometry using the PSF Imagette Photometric Extraction \citep[\texttt{PIPE};][]{Brandeker2024} instead of the aperture photometry produced by \cheops' Data Reduction Pipeline \citep[\texttt{DRP};][]{Hoyer2020}.
We analysed each visit individually using \texttt{pycheops} \citep{Maxted2022} and \texttt{juliet} \citep{Espinoza2019}, a planet modelling tool using nested sampling and implementing \texttt{batman} \citep{Kreidberg2015} to model the transits. We obtained the detrending vectors for each \cheops visit in \texttt{pycheops} via Bayes factor assessments of simultaneously fitting subsets of the detrending vectors with a transit model. We subsequently detrended the visits against the instrumental vectors using linear regressors in \texttt{juliet} to accurately probe the detrending parameter space with the nested sampling algorithm provided in \texttt{juliet}. The detrended vectors for each visit are reported in \autoref{tab:cheops}. We also applied a 3$\sigma$-clipping to the data, removing outliers that were more than 3$\sigma$ away from the mean of the data. As the transit is shallow, no in-transit points were removed by this sigma clipping.

We note that there are an additional four opportunistic \cheops visits available from 2020 September 22 to 2021 September 19. However, since these do not cover any transits, we did not include these in any further analysis.

\begin{table*}
    \centering
    \caption{Summary of the \cheops observations of Gliese 12 and the used detrending vectors.}
    \renewcommand{\arraystretch}{1.4}
    \begin{tabular}{cccccccc}
    \hline
    \hline
        Visit  & Start Date & Duration & Data Points & File Key & Efficiency & Exp Time  & Detrending Vectors\\
        &  (UTC) & (h) & (\#) &  & (\%) & (s) \\
        \hline
        1 & 2023-09-24T00:03:43 & 11.41 & 513 & CH\_PR240007\_TG000101 & 74.9 & 60 & t, x, y$^2$\\
        2 & 2023-10-06T18:19:43 & 12.76 & 481 & CH\_PR240007\_TG000102 & 62.8 & 60 & - \\
        3 & 2023-10-19T16:32:43 & 6.80 & 222 & CH\_PR240012\_TG000701 & 54.3 & 60 & - \\
        4 & 2023-11-14T06:29:42 & 6.92 & 241 & CH\_PR240012\_TG000702 & 57.9 & 60  & - \\
        5 & 2023-11-26T21:38:42 & 7.44 & 248 & CH\_PR240012\_TG000101 & 55.5 & 60 & x, t, cos(2$\phi$)\\
        6 & 2024-08-20T23:26:43 & 6.47 & 279 & CH\_PR240012\_TG000801 & 71.7 & 60 & smear, contam\\
        \hline
    \end{tabular}
    \label{tab:cheops}
\end{table*}

\subsubsection{MuSCAT}
As presented by \citet{Kuzuhara2024}, two transits of Gliese 12\,b have been observed with MuSCAT2 \citep{Narita2019} and two transits with MuSCAT3 \citep{Narita2020}. Directly following the method of \citet{Kuzuhara2024}, we detrended the data using a Gaussian process (GP) regression with a Mat\'ern 3/2 kernel. Similar to our approach with the \textit{CHEOPS} data, we performed this detrending for each visit separately. 

\subsubsection{ASAS-SN}

We retrieved all available data from the All-Sky Automated Survey for Supernovae \citep[ASAS-SN;][]{Shappee2014, Kochanek2017}. Photometry was recomputed via the ASAS-SN web portal, accounting for the proper motion of the star. After excluding poor-quality measurements, the dataset contains 1019 \textit{V}-band observations spanning 2012 January 20 to 2018 November 29 and 3678 \textit{g'}-band measurements spanning 2017 September 21 to 2025 August 26. Outliers were removed using a 5-median absolute deviation (MAD) clipping.

\subsubsection{LCOGT}
We obtained photometric observations with the Las Cumbres Observatory Global Telescope \citep[LCOGT;][]{Brown2013}. The target star was observed using the 0.4\,m and 1.0\,m telescopes in the $V$ and $B$ bands, respectively. For the $B$-band observations, we used an exposure time of 30\,s and collected a total of 570 images between 2023 December 14 and 2025 January 16. For the $V$ band, the exposure time was 300\,s, and we collected 241 images between 2024 August 24 and 2024 December 26. The images were calibrated using the standard LCOGT {\tt BANZAI} pipeline \citep{McCully2018}, and differential photometry was extracted with {\tt AstroImageJ} \citep{Collins2017}.

\subsubsection{TJO}
A long-term photometric campaign of Gliese 12 was carried out using the 0.8\,m Telescopi Joan Or\'o \citep[TJO;][]{Colome2010} at the Montsec Observatory in Lleida, Spain. We obtained a total of 1106 $R$-band images between 2023 May 23 and 2025 January 09. The raw science images were calibrated using the \texttt{ICAT} pipeline \citep{Colome2006}.

\subsubsection{E-EYE}
We collected data with e-EYE\footnote{\url{https://www.e-eye.es/}} (short for `Entre Encinas y Estrellas'), a 16-inch Optimised Dall-Kirkhams (ODK) reflector located at Fregenal de la Sierra in Badajoz, Spain. A total of 351 images were obtained: 175 in the $V$ band and 176 in the $R$ band. The observations span the period from 2023 December 14 to 2025 February 25. Image reduction and differential aperture photometry of the target and several reference stars were performed using the \texttt{LesvePhotometry} package\footnote{\url{http://www.dppobservatory.net/}}.

\subsection{Spectroscopy}

\begin{figure*}
    \centering
	\includegraphics[width=\linewidth]{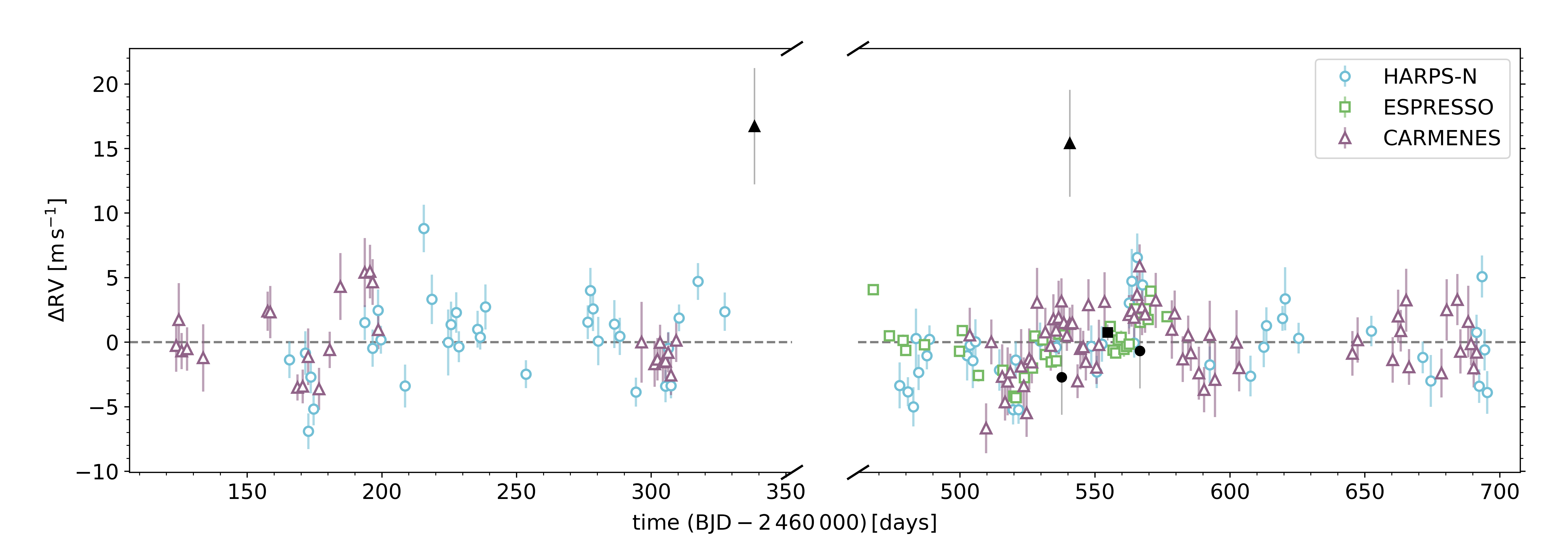}
    \caption{Gliese 12 RVs with offsets removed; HARPS-N, ESPRESSO, and CARMENES data are plotted as blue circles, green squares, and purple triangles, respectively. Uncertainties on the ESPRESSO RVs are smaller than the marker points. Outliers are indicated by the black markers.}
    \label{fig:rvs}
\end{figure*}

In this work, we used three different spectroscopic data sets, taken by HARPS-N, ESPRESSO, and CARMENES. There is also the publicly available data from IRD used by \citet{Kuzuhara2024}. We decided not to use the latter due to the difference in scatter as compared to the more stabilised data used in this work.
Outliers within each used dataset were removed (post-re-extraction in the case of HARPS-N and ESPRESSO) via identifying data points with uncertainties more than 3$\sigma$ from the median uncertainty on an instrument-by-instrument basis.
Summary statistics on the RVs are mentioned in Table~\ref{tab:rvs}, and the full set of RVs used in our analyses are shown in Fig.~\ref{fig:rvs}.

\begin{table}
    \caption{Summary of the RV observations.}
    \label{tab:rvs}
    \renewcommand{\arraystretch}{1.4}
    \begin{tabular}{lccc}
        \hline
        \hline
        Instrument & Number of RVs & Average error & Standard deviation \\
         & &  (m\,s$^{-1}$) & (m\,s$^{-1}$) \\
        \hline
        HARPS-N    & 76 (74) & 1.43 (1.43)  & 2.88 (2.90)  \\
        ESPRESSO   & 35 (34) & 0.35 (0.35) & 1.90 (1.92)  \\
        CARMENES   & 93 (91) & 1.77 (1.77)  & 3.38 (2.47)  \\
        \hline
        \multicolumn{4}{@{}p{\dimexpr\linewidth-2\tabcolsep}@{}}{\small Notes: Values enclosed in brackets are calculated from data after outliers are excluded. For HARPS-N and ESPRESSO, the \sbart values were used.} \\
    \end{tabular}
\end{table}

\subsubsection{HARPS-N}

We obtained 87 observations with the HARPS-N spectrograph \citep{Cosentino2012}, a small subset of which was already published by \citet{Dholakia2024}.
HARPS-N is a high-precision, high-resolution echelle spectrograph installed on the Telescopio Nazionale Galileo (TNG) in La Palma, Spain.
The instrument covers a wavelength range of $383$--$693\,\mathrm{nm}$ and has an average resolving power of $\mathcal{R}=115\,000$.
Observations were taken over four semesters, from 2023 August 09 to 2025 January 28 (BJD 2460165.66 -- BJD 2460704.35), spanning just under a year and a half.
Thirteen observations were obtained via the Guaranteed Time Observation (GTO) programme, with the remaining 63 data points collected as part of the HARPS-N Collaboration under programme A48TAC\_59 (PI: L. Malavolta).

The target was observed with the second fibre pointed at the sky, with observations having an average exposure time of $1800\,\mathrm{s}$. The data were processed with the standard HARPS-N \texttt{DRS} \citep[Data Reduction Software version 3.0.1 -- ][]{Dumusque2021}, with an M5\,V mask used to obtain the cross correlation functions (CCFs). Additionally, the \texttt{DRS} calculated several spectral activity indicators, including the $S$-index and H$\alpha$ metric from the spectra. The median signal-to-noise ratio (S/N) in order 50 is 24.

To ensure homogeneity across all instruments, and accounting for the specific CCF shape of M dwarfs, activity indicators were re-extracted from the CCFs following \citet{Lafarga2020}. An inverted Gaussian was fit to the CCF, from which the contrast, full width at half maximum (FWHM), and bisector inverse slope (BIS) were calculated. To obtain the BIS values, we followed the method outlined by \citet{Queloz2001}, where the top and bottom regions of the CCF are defined as the sections between 60 and 90 per cent, and 10 and 40 per cent, respectively.

As Gliese 12 is an M dwarf, obtaining RVs using the CCF method is challenging due to the large number of spectral lines and their blending. Hence, we re-extracted the RVs using the Semi-Bayesian Approach for RVs with Template-matching\footnote{\url{https://github.com/iastro-pt/sBART}} \citep[\sbart;][]{Silva2022}, which is shown to be especially successful in its application to M dwarfs. Within this process, we re-extracted RVs for different quality checks including all combinations of RV steps (0.1, 0.5, 1.0\,m\,s$^{-1}$), RV limits (200, 500, 1000\,m\,s$^{-1}$), minimum order S/N (1.5, 5, 10), airmasses (1.5, 2.0, 2.2, 2.6), and RV errors (5, 6, 7, 10\,m\,s$^{-1}$), as well as the classical and Laplacian method \sbart applies \citep[see ][ for more detail]{Silva2022}. We selected the re-extracted RVs producing the lowest median error in \sbart. These are for an RV step of 0.1\,m\,s$^{-1}$, RV limit of 500\,m\,s$^{-1}$, minimum order S/N of 1.5, airmass of 1.5, RV errors of 5\,m\,s$^{-1}$, and by using the Laplacian method. We note that the classical method provided very similar results but with slightly higher errors. This removed 11 data points whose spectra did not fulfil \sbart's quality checks and left 76 HARPS-N RVs. The median error of the RVs in \sbart was reduced to 1.43\,m\,s$^{-1}$ as compared to 2.98\,m\,s$^{-1}$ from the \texttt{DRS}. The standard deviation of the RVs became 2.88\,m\,s$^{-1}$ from \sbart compared to 5.10\,m\,s$^{-1}$ from the \texttt{DRS}. 

\subsubsection{ESPRESSO}

Gliese 12 was observed with ESPRESSO \citep{Pepe2014, Pepe2021}, a high-resolution echelle spectrograph installed at the incoherent combined coud\'e facility of the Very Large Telescope (VLT) at the Paranal Observatory, Chile.
ESPRESSO is ultra-stable, covers a wavelength range of $380$--$788\,\mathrm{nm}$, and has an average resolving power of $\mathcal{R}=140\,000$.
A total of 35 observations were taken between 2024 June 06 to 2024 September 23 (BJD 2460467.90 -- BJD 2460576.75) as part of a Cycle P113 programme with ID:113.26RH (PI: Wilson). With an exposure time of 1200\,s, we obtained a median S/N at $\sim$550\,nm of 29.50.
The data were processed through the standard ESPRESSO \texttt{DRS} (version 3.3.1) with an M5\,V mask. Similar to the HARPS-N data, this produced CCFs as well as spectral activity indicators.

We again re-processed the RVs using \sbart for all the combinations of quality checks stated above. We also selected the quality check combination producing RVs with the lowest median error of 0.35\,m\,s$^{-1}$ compared to a median error of 0.70\,m\,s$^{-1}$ from the \texttt{DRS}. This resulted in a standard deviation of 1.90\,m\,s$^{-1}$ for \sbart and 2.14\,m\,s$^{-1}$ for the \texttt{DRS}. For ESPRESSO, this is the case for an RV step of 0.1\,m\,s$^{-1}$, RV limit of 200\,m\,s$^{-1}$, minimum order S/N of 1.5, airmass of 1.5 and RV error of 7\,m\,s$^{-1}$, as well as applying the Laplacian method in \sbart. This method kept in all 35 RV observations taken by ESPRESSO. The RVs utilised in our analyses were those extracted using \sbart.

\subsubsection{CARMENES}

Gliese 12 was observed with the CARMENES spectrograph installed at the 3.5\,m telescope of Calar Alto Observatory in Almer\'{i}a, Spain. CARMENES has two channels, one operating at visible wavelengths (VIS, spectral range 0.52--0.96\,$\mu$m) and one operating at near-infrared wavelengths (NIR, spectral range 0.96--1.71\,$\mu$m). The average spectral resolution of the VIS channel is $\mathcal{R} = 94\,600$ and the resolution of the NIR channel is $\mathcal{R} = 80\,400$ \citep{Quirrenbach2014, Quirrenbach2018}. 

A total of 93 RV measurements were collected from 2023 June 28 to 2025 January 15 (BJD 2460123.59 -- BJD 2460691.34), and part of the data collected in 2023 was presented by \citet{Kuzuhara2024}. The spectra were acquired with an exposure time of 1800\,s. Data reduction was performed using the \texttt{CARACAL} pipeline \citep{Caballero2016}, which applies standard corrections for bias, flat field, and cosmic rays, and extracts spectra using the {\tt FOX} optimal extraction algorithm \citep{Zechmeister2014}. Wavelength calibration was carried out following the procedure described by \citet{Bauer2015}. RV measurements were obtained with the template-matching algorithm \texttt{SERVAL}\footnote{\url{https://github.com/mzechmeister/serval}} \citep{Zechmeister2018}. The RV measurements were corrected for other effects such as barycentric motion, secular acceleration, instrumental drifts, and nightly zero-points \citep[e.g.,][]{Trifonov2018,TalOr2018}.

The 93 spectra have a median S/N of 90 measured at the spectral order centred at $746\, \mathrm{nm}$ (minimum $\mathrm{S/N} = 27$, maximum $\mathrm{S/N} = 118$) in the VIS channel. The median value of the RV uncertainties are 1.76\,m\,s$^{-1}$ for the VIS channel, and 6.24\,m\,s$^{-1}$ for the NIR channel. In this work, we used only the VIS channel RVs, as they present significantly lower scatter than the NIR velocities.

\section{Host Star characterisation}
\label{sec:star}

Gliese 12 is a nearby M dwarf with sub-solar metallicity, likely belonging to the Galactic thin disc; see \citet{Cortes-Contreras2024} and \citet{Dholakia2024} for a discussion of the kinematics.

\subsection{Stellar parameters}

It was previously characterised by \citet{Dholakia2024} and \citet{Kuzuhara2024}. Using the CARMENES data, we determined the stellar atmospheric parameters, namely $T_{\rm eff}$, $\log{g}$, and [Fe/H], on the CARMENES template spectrum corrected for telluric absorption \citep{Nagel2023} with the code {\tt SteParSyn}\footnote{\url{https://github.com/hmtabernero/SteParSyn/}} \citep{Tabernero2022a}  using the line list and model grid described by \cite{Marfil2021}. We set the total line broadening to account for both the macroturbulence and the projected rotational velocity of the star \citep[$v_{\rm broad}$, see][]{Tabernero2022a} to 2\,km\,s$^{-1}$. This is supported by a model-independent determination using the projected rotational velocity ($v\sin{i}$) of \citet{Reiners2018}. The luminosity was derived from the integration of the spectral energy distribution following \citet{Cifuentes2020} with updated photometry from $B$ to $W4$ and \textit{Gaia} DR3 parallax \citep{GaiaDR3}. The stellar radius follows the Stefan-Boltzmann law and the stellar mass from the linear mass-radius relation given by \citet{Schweitzer2019}. Table~\ref{tab:stellar_params} summarises our derived parameters based on the CARMENES data, which are easily consistent with the errors of the two previous characterisations.

The stellar elemental abundances for Gliese 12 were derived following the methodology of \citet{Gromek2025}, based on the work of \citet{Hejazi2023}, and are also summarised in Table~\ref{tab:stellar_params}. We performed a spectral synthesis analysis using the 1D telluric-corrected CARMENES template spectrum. Synthetic spectra were generated with MARCS stellar atmosphere models \citep{Gustafsson2008} and the \texttt{Turbospectrum} radiative transfer code \citep{Alvarez1998, Plez2012}, implemented via modified \textsc{iSpec} routines \citep{Blanco-Cuaresma2014, Blanco-Cuaresma2019} using solar abundances from \citet{Asplund2009}. The stellar parameters adopted to generate the model spectra were $T_\mathrm{eff} = 3328 \pm 78\,\mathrm{K}$, $\log{g} = 5.00 \pm 0.05\,\mathrm{dex}$, $[\mathrm{m/H}] = -0.30 \pm 0.11\,\mathrm{dex}$, $v_\mathrm{mac} = 5.0 \pm 1.0\,\mathrm{km\,s^{-1}}$, and $v_\mathrm{mic} = 0.5 \pm 0.5\,\mathrm{km\,s^{-1}}$. Of these parameters, $T_\mathrm{eff}$ and $[\mathrm{m/H}]$ were taken directly from the results of the \texttt{SteParSyn} analysis, with $\log{g}$ calculated using the new mass and radius values. The macroturbulence and microturbulence velocities were determined by performing a $\chi^2$-minimisation over molecular OH lines, which are especially sensitive to these broadening parameters \citep{Souto2017, Hejazi2023}. Candidate absorption lines were selected from the normalised spectrum and cross-referenced with atomic and molecular features in the VALD line list \citep{Kupka2011}. The line list was further refined via visual inspection to exclude blended or contaminated features. The oxygen abundance was calculated using OH molecular features in place of OI lines.

For each spectral line, we generated a grid of synthetic spectra by varying the elemental abundance, $[\mathrm{X/H}]$, between $-0.75$ and $+0.75\,\mathrm{dex}$ in steps of $0.25\,\mathrm{dex}$, followed by interpolating the model spectra to a finer grid with $0.015\,\mathrm{dex}$ resolution. Best-fit abundances were determined for each line via $\chi^2$-minimisation between the model and observed spectrum within a fitting window. The final abundances for each element are calculated via a weighted average of the individual abundances for each spectral line, with the weighting for each line equal to the root mean square error between the best-fit model and the observed line in each line region, divided by the line depth. The uncertainties due to scatter ($\sigma_\mathrm{rms}$) were calculated as the standard deviation of the line-by-line abundance distribution divided by $\sqrt{N}$, where $N$ is the number of lines used per element. Systematic uncertainties due to $T_\mathrm{eff}$, $[\mathrm{m/H}]$, $\log{g}$, $v_\mathrm{mac}$, and $v_\mathrm{mic}$ were estimated by independently resampling each stellar parameter from its Gaussian uncertainty distribution and repeating the full analysis over 15 iterations \citep{Hejazi2023}. The total uncertainty for each elemental abundance was computed by summing the random and systematic components in quadrature. Using the final abundances, we recomputed the global metallicity and $\alpha$-enhancement of Gliese 12 following the prescription of \citet{Hinkel2022}. By adding the number ratios of the available metals (O, Na, Mg, Si, K, Ca, Ti, Cr, Mn and Fe), we find an overall metallicity of $[\mathrm{m/H}] = -0.18 \pm 0.05\,\mathrm{dex}$, consistent with the metallicity reported by \citet{Newton2014}, who determined an [Fe/H] of $-0.17 \pm 0.13\,\mathrm{dex}$.

\begin{table}
        \centering
        \caption{Gliese 12 stellar parameters used in this work.}
        \renewcommand{\arraystretch}{1.4}
        \begin{tabular}{lccc}
            \hline
            \hline
             Parameter & Unit & Value & Source \\
             \hline 
             $T_\mathrm{eff}$ & (K) & $3328\pm78$ & (a) \\
             $M_\star$ & (M$_\odot$) & $0.255 \pm 0.013$ & (a) \\
             $R_\star$ & (R$_\odot$) & $0.265\pm0.012$ & (a) \\
             $\rho_\star$ & ($\mathrm{g\,cm^{-3}}$) & $19.3^{+3.1}_{-2.6}$ & (a) \\
             $\log{g}$ & (dex) & $5.25\pm0.09$ & (a) \\
             Age & (Gyr) & $7.0_{-2.2}^{+2.8}$ & (b) \\
             Distance & (pc) & $12.1616_{-0.0053}^{+0.0051}$ & (c) \\
             $[\mathrm{m/H}]$ & (dex) & $-0.18 \pm 0.05$ & (a) \\
             $[\mathrm{Fe/H}]$ & (dex) & $-0.24 \pm 0.09$ & (a) \\
             $[\mathrm{O/H}]$ & (dex) & $-0.16 \pm 0.05$ & (a) \\
             $[\mathrm{Na/H}]$ & (dex) & $-0.20 \pm 0.09$ & (a) \\
             $[\mathrm{Mg/H}]$ & (dex) & $-0.41 \pm 0.06$ & (a) \\
             $[\mathrm{Si/H}]$ & (dex) & $-0.14 \pm 0.16$ & (a) \\
             $[\mathrm{K/H}]$ & (dex) & $-0.25 \pm 0.09$ & (a) \\
             $[\mathrm{Ca/H}]$ & (dex) & $-0.25 \pm 0.09$ & (a) \\
             $[\mathrm{Ti/H}]$ & (dex) & $-0.33 \pm 0.08$ & (a) \\
             $[\mathrm{Cr/H}]$ & (dex) & $-0.20 \pm 0.05$ & (a) \\
             $[\mathrm{Mn/H}]$ & (dex) & $-0.16 \pm 0.08$ & (a) \\
             $P_\mathrm{rot}$ & (d) & $\sim85$ & (a), (d) \\
             $\langle \log{R'_\mathrm{HK}} \rangle$ & --- & $-5.62^{+0.10}_{-0.14}$ & (a) \\ 
             \hline
             \multicolumn{4}{@{}p{\dimexpr0.85\linewidth-2\tabcolsep}@{}}{Notes: (a) This work, (b) \citet{Dholakia2024}, (c) \citet{Bailer-Jones2021}, (d) \citet{Kuzuhara2024}.}
        \end{tabular}
        \label{tab:stellar_params}
\end{table}

\subsection{Stellar variability}
\label{sec:starvar}

Gliese 12 is overall a very quiet star as evidenced by its low $\langle \log{R'_\mathrm{HK}} \rangle$ value, a low X-ray flux, and low variability in the photometry.
We recalculated the $\langle \log{R'_\mathrm{HK}} \rangle$ from the HARPS-N spectra and following the expressions of \citet{Astudillo-Defru2017}. We found that over the timespan of the data, $\langle \log{R'_\mathrm{HK}} \rangle = -5.62$. Using the estimated relation between $\langle \log{R'_\mathrm{HK}} \rangle$ and the rotation period from \citet{Astudillo-Defru2017}, the median rotation period is predicted to be $\sim$75\,d. 

We attempted to determine the stellar rotation period of Gliese 12 using long-term, ground-based photometric data from ASAS-SN, LCOGT, TJO, and E-EYE (see Fig. \ref{fig:longterm-photometry}). Each dataset was analysed independently — except for the E-EYE data, which included simultaneous $V$ and $R$ observations on the same nights — to assess whether they yielded consistent estimates of $P_\mathrm{rot}$. To model the photometric time series, we fitted a linear function to account for long-term trends, alongside a sinusoidal function:
\begin{equation}
f(t) = z + mt + A \sin\left( \frac{2\pi t}{P_\mathrm{rot}} + \phi \right).
\label{eq:photometry_equation}
\end{equation}
Here, $t$ denotes the epoch of the photometric measurements. The free parameters include $z$ and $m$, the coefficients of the linear trend, and the sinusoidal components: amplitude $A$, rotation period $P_\mathrm{rot}$, and phase $\phi$. To account for correlated noise present in the data, we employed GPs \citep[e.g.,][]{Rasmussen2006,Gibson2012}, specifically using an exponential kernel implemented in \texttt{Celerite}\footnote{\url{https://celerite.readthedocs.io/en/stable/}} \citep{ForemanMackey2017}. We ran \texttt{PyDE}\footnote{\url{https://github.com/hpparvi/PyDE}} to find optimal initial solutions, followed by a Markov chain Monte Carlo (MCMC) analysis with \texttt{emcee} \citep{ForemanMackey2013} to determine posterior distributions (50 chains, $2\times 10^4$ iterations with a thin factor of 100). Final parameter values were derived from these, using the median and 1$\sigma$ percentiles.

The derived $P_\mathrm{rot}$ values are listed in Table~\ref{tab:prot_values}. Our results indicate a relatively long stellar rotation period, with most estimates falling in the 80–90 day range. The most precise determination comes from the ASAS-SN $g'$ data, yielding $P_\mathrm{rot} = 84.9_{-0.3}^{+0.4}\,\mathrm{d}$, consistent with the rotation period obtained by \citet{Kuzuhara2024} with ASAS-SN data. Rotation periods derived from $V$ band photometry are considerably less well determined.

\begin{figure}
    \centering
    \includegraphics[width=0.95\linewidth]{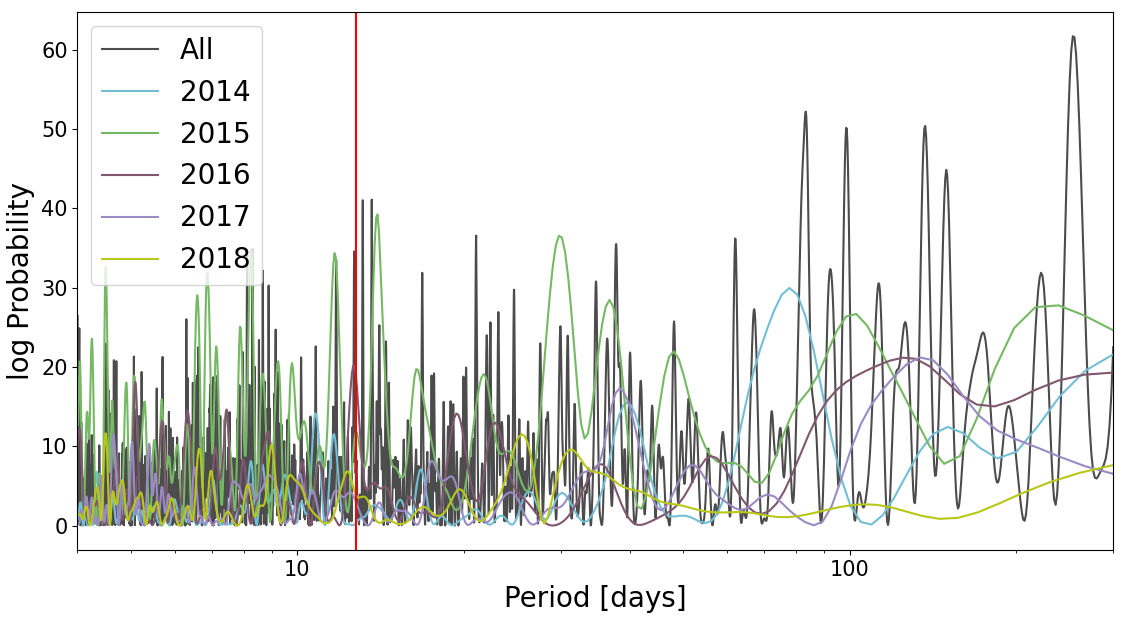}

    \includegraphics[width=0.95\linewidth]{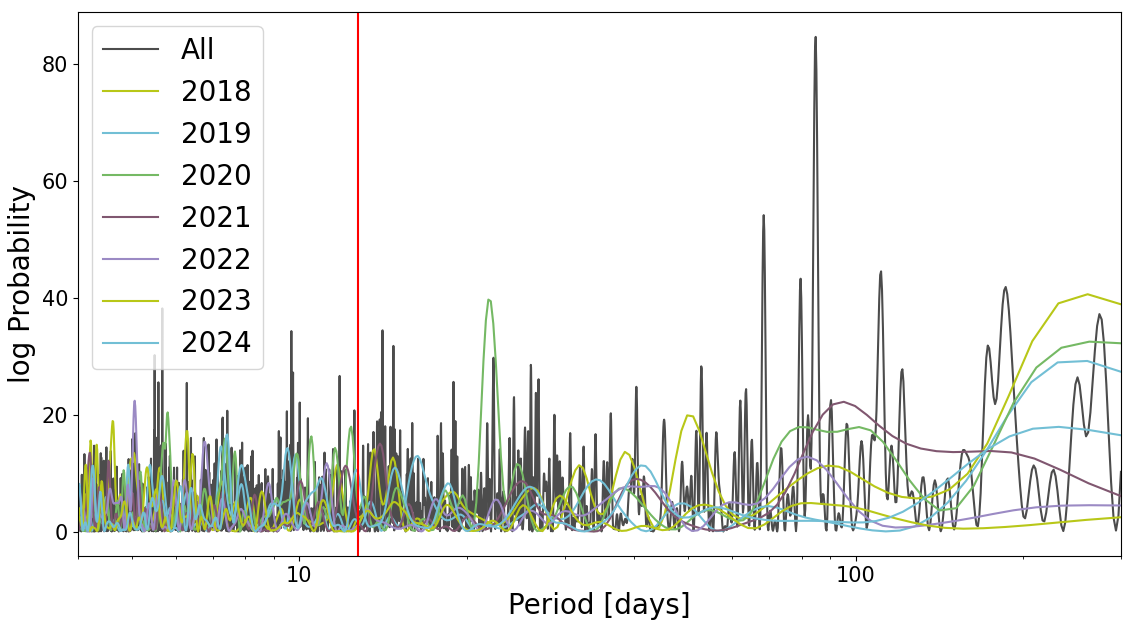}
    \caption{BGLS periodogram of the ASAS-SN $V$ band (top) and $g'$ band (bottom) data of Gliese 12, of the full data and per season. The red vertical line indicates the planet period.}
    \label{fig:asas_sn_bgls}
\end{figure}

Applying a BGLS periodogram \citep{Mortier2015} to the ASAS-SN data and splitting the data into the different observing seasons, we found that, unsurprisingly, the variability is season-dependent with several years showing no variability around 80 days across both photometric bands. The periodogram is shown in Fig.~\ref{fig:asas_sn_bgls}. It confirms that this star is relatively quiet and that the variability pattern is not clear-cut. We caution that this may contribute to having the rotation periods better or less well determined above as it will partly depend on the seasons when the photometry was taken.

\begin{table} 
        \centering
        \caption{Stellar rotation estimates from long-term photometry.}
        \renewcommand{\arraystretch}{1.4}
        \begin{tabular}{lcc}
            \hline
            \hline
             Facility & Photometric band & Value \\
             \hline 
             ASAS-SN & $V$ & $P_\mathrm{rot} = 80.3_{-59.8}^{+23.9}$\,d \\
             ASAS-SN & $g'$ & $P_\mathrm{rot} = 84.9_{-0.3}^{+0.4}$\,d \\
             LCOGT & $B$ & $P_\mathrm{rot} = 92.7_{-2.9}^{+4.8}$\,d \\
             LCOGT & $V$ & $P_\mathrm{rot} = 86.6_{-32.5}^{+22.3}$\,d \\
             TJO & $R$ & $P_\mathrm{rot} = 77.0_{-1.9}^{+1.9}$\,d \\
             E-EYE & $V$, $R$ & $P_\mathrm{rot} = 81.8_{-3.2}^{+3.3}$\,d \\
             \hline
        \end{tabular}
        \label{tab:prot_values}
\end{table}

Finally, we looked at the discrete correlation function (DCF) between H$\alpha$ and RVs. Since these are measured from the same spectra and available for all our spectroscopic data sets, they could be used in a multi-dimensional GP regression fit. We applied the algorithm from \citet{Edelson1988} to compute the DCF. The resulting analysis shows that H$\alpha$ and RVs are mildly anti-correlated and that there is a small timeshift of about 10\,d. This can again point to a rotation period of 80\,d where any time shift expresses itself at about one eighth of the rotation period \citep{CollierCameron2019,Mortier2022}. Perhaps more intriguingly, the data become mildly correlated after 50--70\,d (depending on the direction of the time shift). This points to a rotational correlation timescale between H$\alpha$ and RV of 120\,d. While this is unlikely to be the rotational period considering the alternative measurements, it appears to be the timescale manifested in the data that we have. As seen later in Section \ref{sec:RVonly}, several values are identified for the GP rotational periodicity parameter when applying a wide GP prior, including 80 and 120 days. The mass of the planet is not affected by this choice.

\section{Photometry modelling}
\label{sec:Photonly}

\begin{figure*}
    \centering
    \includegraphics[width=0.79\linewidth]{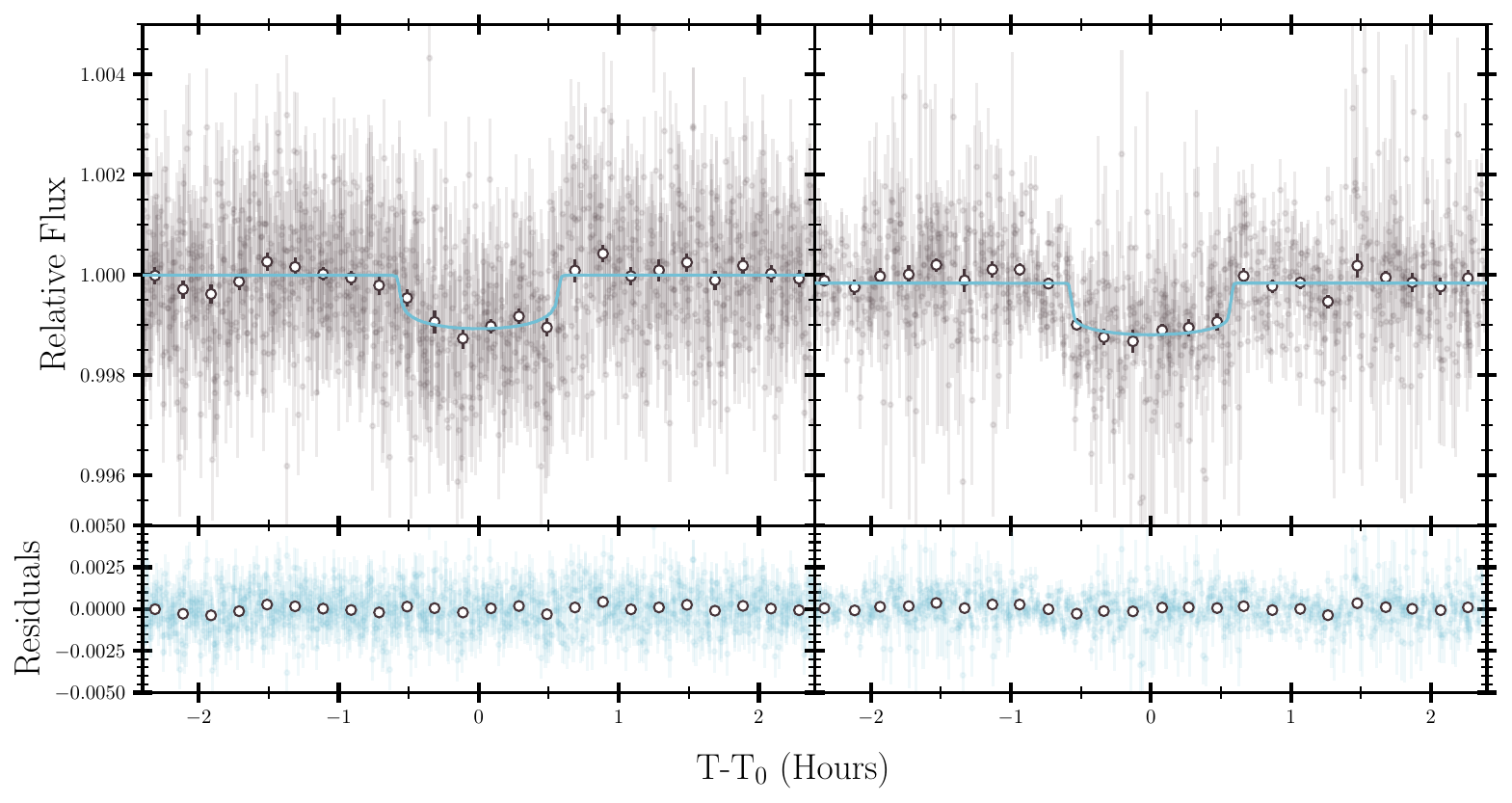}
    \includegraphics[width=0.79\linewidth]{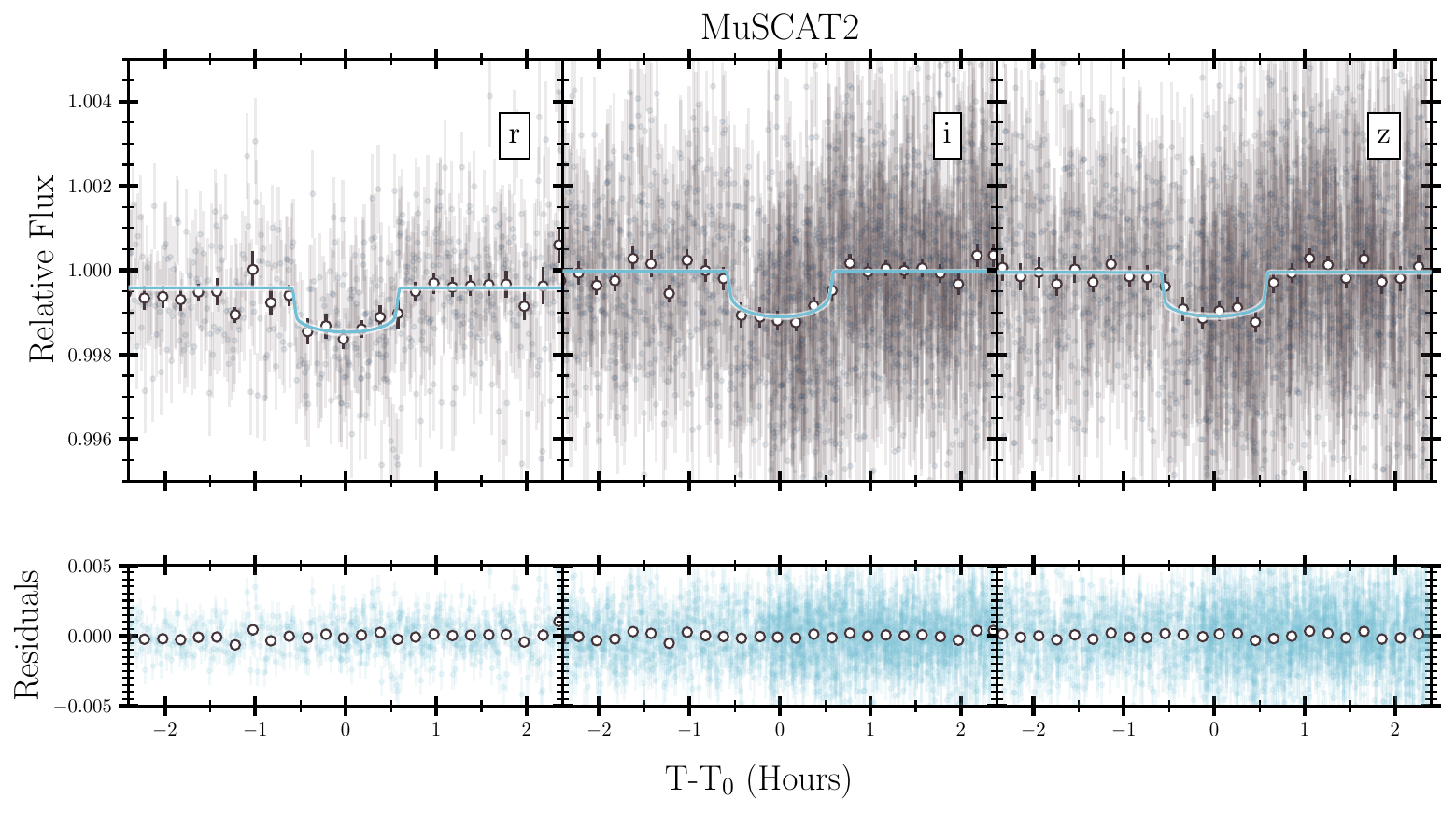}
    \includegraphics[width=0.79\linewidth]{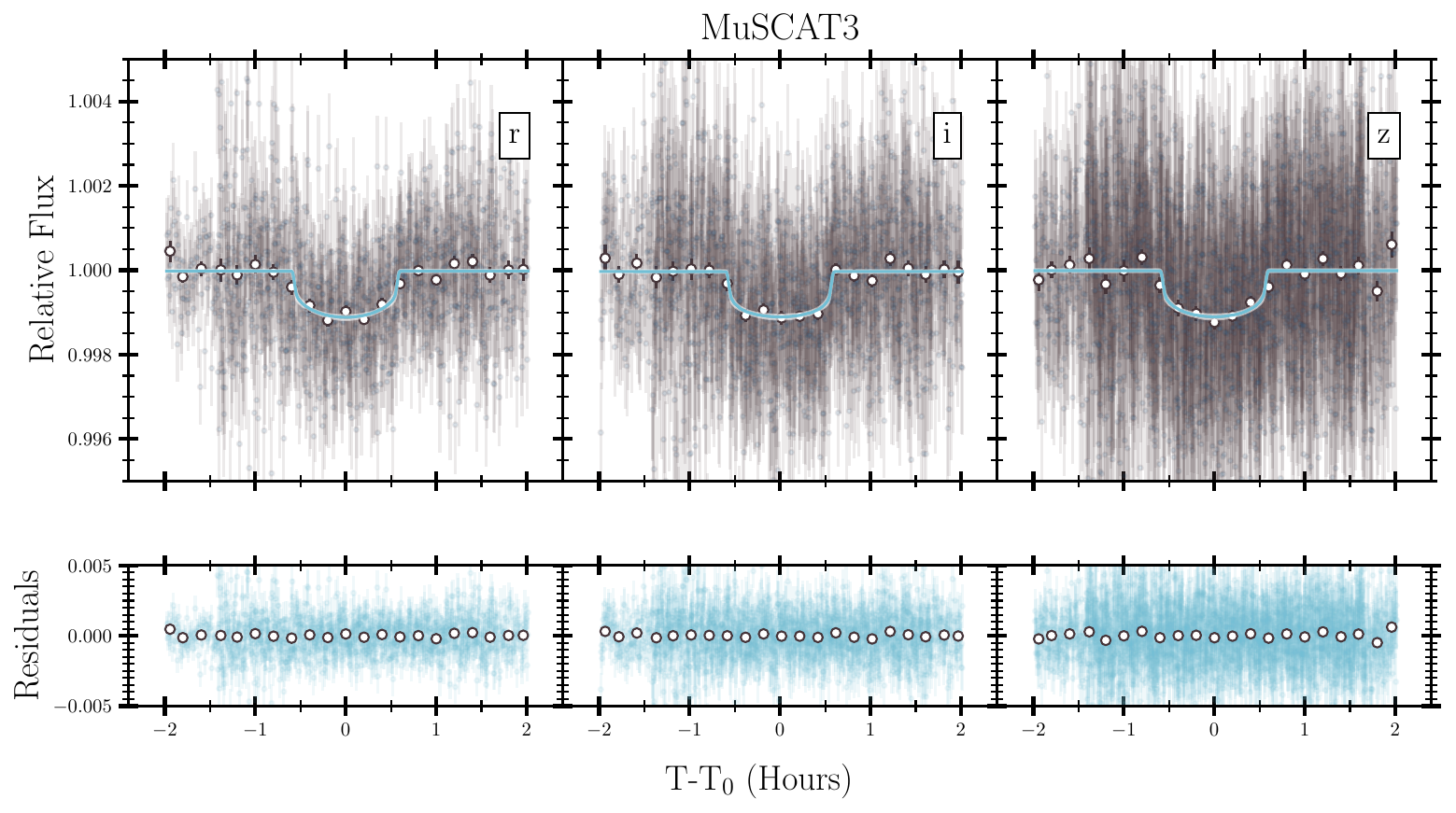}
    \caption{Detrended and phase-folded transits of Gliese 12\,b with the best-fitted model from \texttt{juliet} plotted in blue as well as the residuals on the bottom. Top left: \tess; top right: \cheops, middle: MuSCAT2; bottom: MuSCAT3.}
    \label{fig:transits}
\end{figure*}

We fit the available photometry from \tess, \cheops, and MuSCAT2/3 jointly in \texttt{juliet} \citep{Espinoza2019}. We used the six detrended \cheops visits that contain a transit of Gliese 12\,b as well as the five \tess sectors, and four detrended MuSCAT2/3 observations across \textit{r}, \textit{i}, and \textit{z} bands. In order to detrend the \tess data, we included a GP with the approximate Matern kernel implemented into juliet with celerite \citep{celerite} (fitting for the amplitude of the GP, GP$_\sigma$, and the time-scale of the GP, GP$_\rho$). We fit for:
\begin{description}
    \item the period, $P$;
    \item the time-of-transit centre, $T_0$;
    \item the planet-to-star radius ratio, $R_\mathrm{pl}/R_\star$;
    \item the impact parameter, $b$;
    \item the stellar density, $\rho$;
    \item the limb darkening coefficients parametrised following \citet{Kipping2013}, for \tess, \cheops, and MuSCAT2/3 \textit{r}, \textit{i}, and \textit{z} bands, respectively;
    \item the jitter, for \tess, \cheops, and MuSCAT2/3 \textit{r}, \textit{i}, and \textit{z} bands, respectively;
    \item the offset relative flux, for \tess, \cheops, and MuSCAT2/3 \textit{r}, \textit{i}, and \textit{z} bands, respectively.
\end{description}

Due to the isolated field, we fixed the dilution factor for \tess, \cheops, and MuSCAT2/3 \textit{r}, \textit{i}, and \textit{z} bands to 1. We used the values for the period and time-of-transit centre derived from \citet{Kuzuhara2024} as priors in a normal distribution but inflated the width to be 0.1. We also set a normal distribution for the stellar density with the stellar values obtained from \citet{Kuzuhara2024}. We let the other parameters vary uniformly. All the prior distributions are listed in Tables \ref{tab:fit_priors_planet} and \ref{tab:fit_priors_starinst}.

Finally, we performed two different fits, one assuming a circular orbit and one with free eccentricity and argument of periastron. Both fits resulted in higher precisions on the period and time-of-transit centre as compared to \citet{Dholakia2024} and \citet{Kuzuhara2024}. Furthermore, the additional \tess sector and \cheops visit as well as the re-extraction of the \cheops data, and joint fitting with the MuSCAT2/3 photometry allows us to derive the planet radius from the fit more precisely. From the circular fit, we found a radius of $0.96\pm0.03\,\mathrm{R_\oplus}$ while the eccentric solution gave a radius of $0.90\pm0.04\,\mathrm{R_\oplus}$ with an eccentricity of $0.39^{+0.17}_{-0.11}$.

We adopted the eccentric fit as our final result. We show the phase-folded transits with the best-fit model from \texttt{juliet} in \autoref{fig:transits}. Final extracted values for the orbital and planetary parameters are listed in Table \ref{tab:fit_results_planet} and the stellar and instrumental parameters are listed in Table \ref{tab:fit_results_starinst}.

\begin{table*}
        \centering
        \caption{Gliese 12\,b orbital and planetary parameters for the three different fitting procedures. The fitted parameters are above the horizontal line and the derived parameters below. Exceptions are mentioned in the notes below the table.}
        \renewcommand{\arraystretch}{1.4}
        \begin{tabular}{lcccc}
        \hline
        \hline
             Parameter & Unit & Photometry-only & Informed RV fit & Joint fit\\
             \hline
             $P$ & (d) & $12.761417^{+0.000047}_{-0.000037}$ & $12.761421 \pm 0.000047$ & $12.761418_{-0.000055}^{+0.000060}$ \\
             $T_0$ & (BJD) & $2459497.1855^{+0.0020}_{-0.0026}$ & $2460492.5745_{-0.0043}^{+0.0045}$ & $2460033.16351_{-0.00088}^{+0.00087}$ \\
             $R_\mathrm{p}/R_\star$ & --- & $0.03080^{+0.00084}_{-0.00073}$ & --- & $0.0323_{-0.0014}^{+0.0015}$ \\
             $b$ & --- & $0.51^{+0.12}_{-0.12}$ & --- & $0.75_{-0.15}^{+0.09}$ \\
             $\rho_\star$ & (g\,cm$^{-3}$) & $19.211^{+1.183}_{- 1.198}$ & --- & $19.518^{+2.812}_{-2.766}$ \\
             $K$ & (m\,s$^{-1}$) & --- & $0.70_{-0.20}^{+0.19}$ & $0.67_{-0.21}^{+0.21}$ \\
             $\sqrt{e}\sin{\omega}$ & --- & --- & $-0.12_{-0.33}^{+0.38}$ & $-0.09_{-0.33}^{+0.40}$ \\
             $\sqrt{e}\cos{\omega}$ & --- & --- & $0.35_{-0.20}^{+0.12}$ & $0.35_{-0.24}^{+0.12}$ \\
             $e$ & --- & $0.39^{+0.17}_{-0.11}$ & $0.24_{-0.12}^{+0.15}\;(<0.53)$ & $0.24_{-0.13}^{+0.13}\;(<0.46)$ \\
             $\omega$ & (deg) & $143^{+28}_{-39}$ & $-18_{-41}^{+61}$ & $-16_{-41}^{+68}$ \\
             \hline
             $a/R_\star$ & --- & $54.9^{+1.1}_{-1.2}$ & --- & $55.2_{-2.7}^{+2.5}$\\
             $i$ & (deg) & $89.47\pm0.13$ & --- & $89.25_{-0.09}^{+0.08}$\\
             $M_\mathrm{p}$ & ($\mathrm{M_\oplus}$) & --- & $0.98_{-0.28}^{+0.27}$ & $0.95_{-0.30}^{+0.29}$ \\
             $R_\mathrm{p}$ & ($\mathrm{R_\oplus}$) & $0.90^{+0.04}_{-0.03}$  & --- & $0.93^{+0.06}_{-0.06}$ \\
             $\rho_\mathrm{p}$ & (g\,cm$^{-3}$) & \multicolumn{2}{c}{$7.0^{+2.3}_{-2.1}$} & $6.4 \pm 2.4$ \\
             \hline
            \multicolumn{5}{@{}p{\dimexpr0.65\linewidth-2\tabcolsep}@{}}{\small Notes: For the photometric fit, $e$ and $\omega$ are fitted directly; in the informed RV and the joint fit, they are derived. Numbers in brackets indicate the $2\sigma$ upper limit for the value.} \\
        \end{tabular}
        \label{tab:fit_results_planet}
\end{table*}

\section{RV modelling}
\label{sec:RVonly}

We fitted the RVs (HARPS-N \sbart, ESPRESSO \sbart, and CARMENES \texttt{SERVAL}) using the \pyaneti package\footnote{\url{https://github.com/oscaribv/pyaneti}}, presented by \citet{Barragan2019, Barragan2022}.
Within a periodogram of the RV data (see Fig.~\ref{fig:RV_periodograms}), the signal of the planet is not apparent to any significance. However, since the orbital period and time-of-transit are known from photometric modelling, we used these results to inform the corresponding priors for this RV fit. Despite the small value of \logrhk, as shown in Table~\ref{tab:stellar_params}, there are also signs in the periodogram of possibly stellar-related signals in the data that manifest as peaks in the long-period regime (see Figure~\ref{fig:RV_periodograms}).

We used \texttt{pyaneti} as it is one of few packages that can model the RVs with a single GP, but it can also do multi-dimensional GP fitting, where multiple data sets are co-modelled with one common GP \citep{Rajpaul2015,Barragan2022}.
We used both functionalities when modelling the RVs, and also tried fitting without any GP at all.
In all cases where a GP was used, we employed the quasi-periodic kernel and fit for the GP amplitudes ($A_0$ for the single GP and adding in $A_1$ and $A_2$ for the multi-GP), the activity decay timescale ($\lambda_\mathrm{e}$), the inverse harmonic complexity ($\lambda_\mathrm{p}$), and the stellar rotation period ($P_\mathrm{rot,GP}$); see \citet{Rajpaul2015} and \citet{Barragan2022} for a more detailed description of the kernel.

In all fits, the eccentricity, $e$, and argument of periastron, $\omega$, were reparameterised as $\sqrt{e}\sin{\omega}$ and $\sqrt{e}\cos{\omega}$, as in \citet{Eastman2013}. As aforementioned, the orbital period, $P$, and transit time, $T_0$, were fit with Gaussian priors, informed by the results of the photometry-only fit; the prior on $T_0$ was adjusted (accounting for the uncertainty in $P$) to ensure that it fell near the midpoint of the RV data. Finally, offsets and jitter corresponding to the individual instruments were also fitted as free parameters in the models.

A multitude of runs were performed, including circular and eccentric models; no GP, single GP and multi-dimensional GPs. For the latter, various combinations of parameters (i.e., FWHM, BIS, and H$\alpha$) were utilised in combination with the RVs. The rotation period prior was varied, allowing for a wide prior, a prior centred around 80\,d motivated by the photometry, and a prior centred around 120\,d motivated by the results of the wide prior and the coherence timescale between the RVs and the H$\alpha$ indicator from the DCF analysis. All results constrain the planet parameters similarly and they are all consistent within less than $0.5\sigma$. Several results are virtually indistinguishable for the planet parameters, but the BIC values show preference for the model with a GP applied to the RVs with $\Delta \mathrm{BIC} = 77.9$ compared to the next-best model (no GP). We decided to adopt the results achieved from the model using a standard GP applied solely to the RV data, as the BIC values show preference for this model. All prior distributions of the adopted fit can be found in Tables~\ref{tab:fit_priors_planet} and \ref{tab:fit_priors_starinst}.

\begin{figure}
    \centering
    \includegraphics[width=\linewidth]{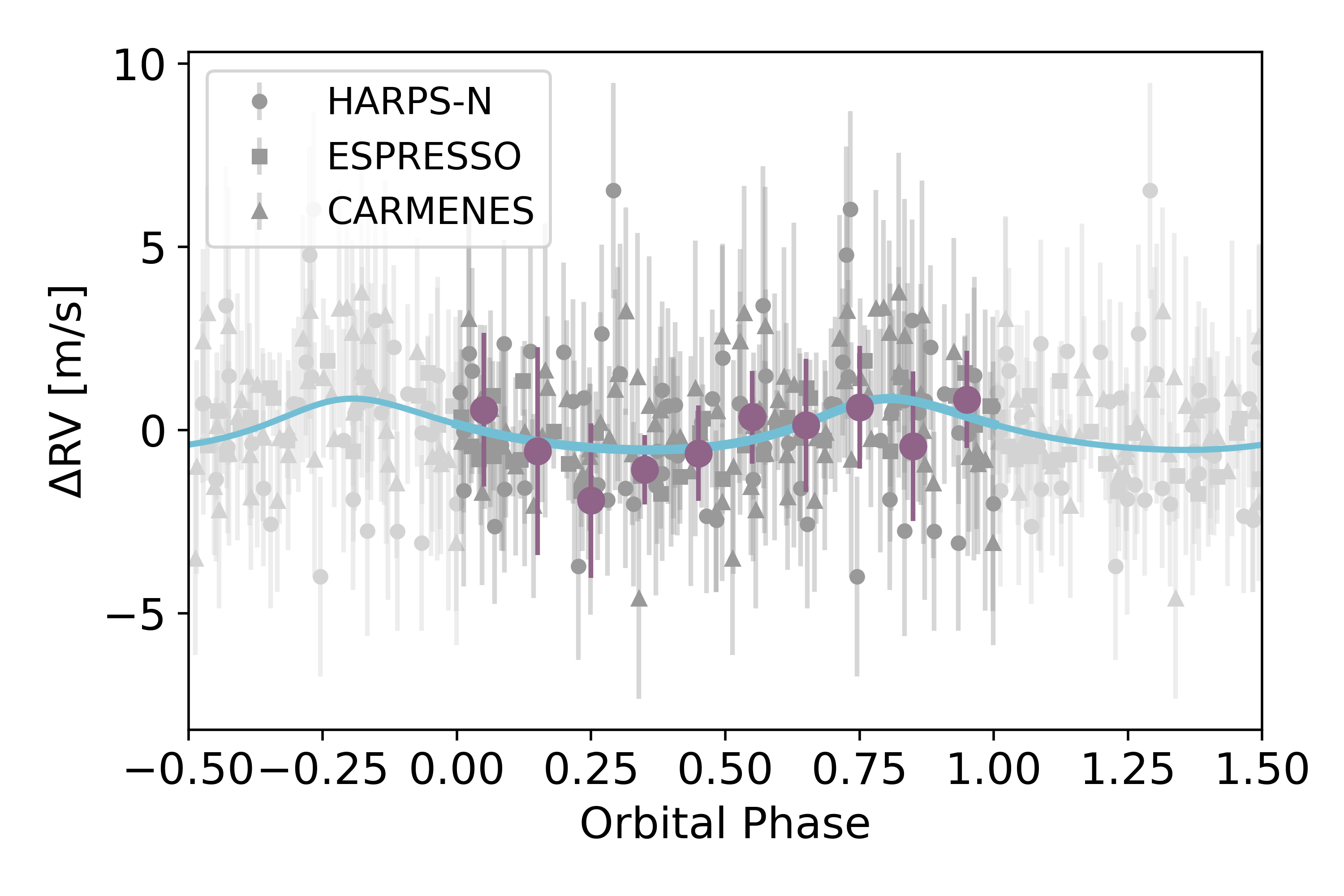}
    \caption{Phase-folded RVs from the best-fit informed RV model with GPs removed. Binned data shown by the large, purple points (10 bins). Phase-folded on the period of the planet (12.761421\,d).}
    \label{fig:phase_folded_RV}
\end{figure}

\begin{figure}
    \centering
    \includegraphics[width=\linewidth]{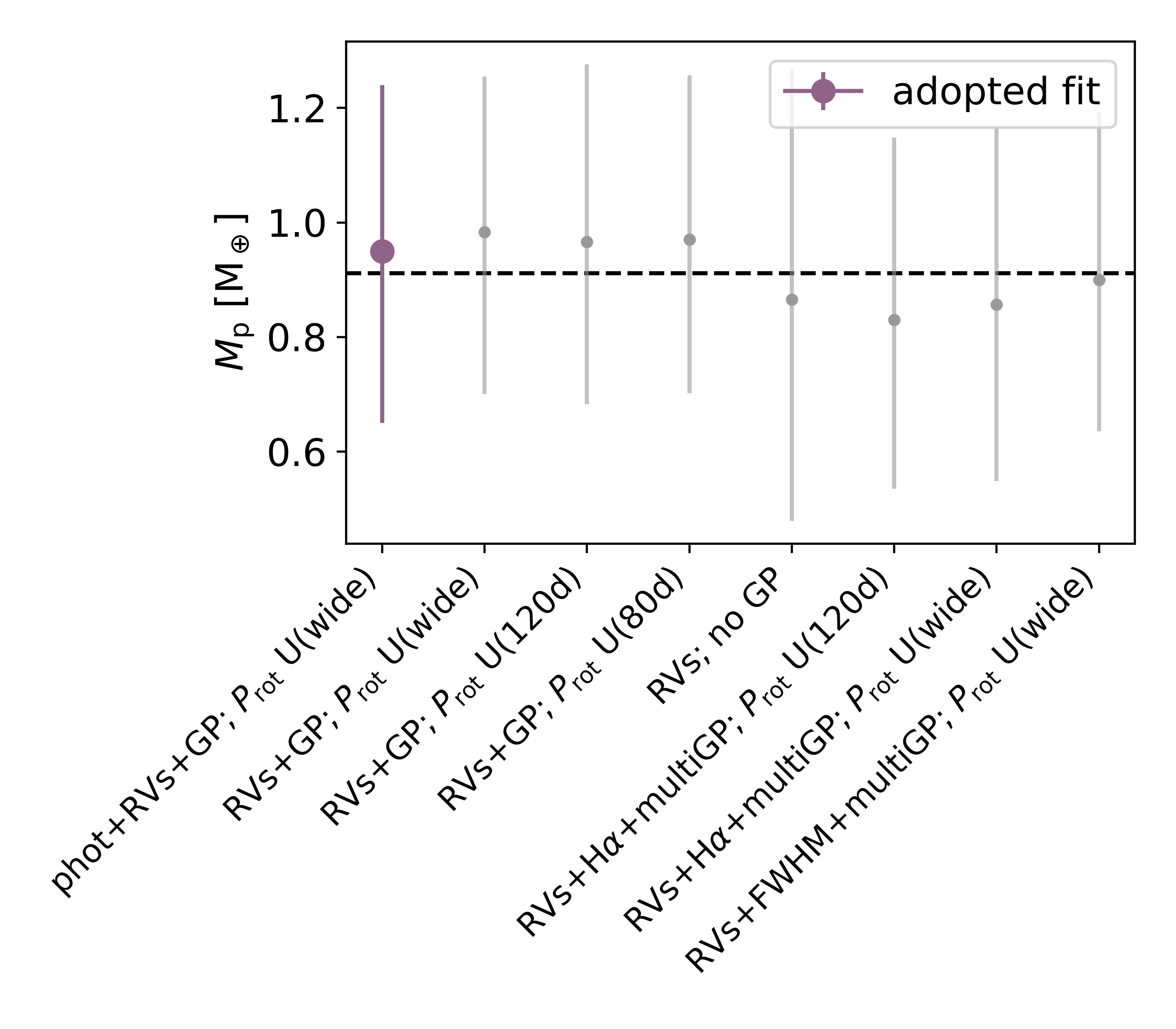}
    \caption{A summary of the variety of models fitted to the RV data. The purple point highlights the adopted fit. Here, $\mathcal{U}$(wide) indicates wide uniform priors for the GP period (as in Table~\ref{tab:fit_priors_starinst}), and $\mathcal{U}$(value) indicates uniform priors defined with 15\,d either side of the specified value. The dashed, black line indicates the weighted mean of all the mass values.}
    \label{fig:model_comp}
\end{figure}

We measured an RV semi-amplitude of $K=0.70\,\mathrm{m\,s^{-1}}$ with a $3.6\sigma$ significance. The phase-folded model and data are presented in Fig.~\ref{fig:phase_folded_RV}.
We thus found that Gliese 12\,b has a mass of $0.98^{+0.27}_{-0.28}\,\mathrm{M_\oplus}$, and an eccentricity of $0.24^{+0.15}_{-0.12}$ constrained only at the 2$\sigma$ level. With the current dataset, we were not able to more-precisely constrain the eccentricity.
Upon checking all the individual results, high consistency on the planet parameters was found between runs where parameters agreed well within 0.5$\sigma$. For some runs, this is shown in Fig.~\ref{fig:model_comp}.

It is recognised that the period of the GP does not always converge at the value of the rotational period that was identified via photometry (85\,d). Additionally, constraining the GP period of around 85\,d resulted in a poorer fit in comparison to a wider prior. Given that $\langle\log{R'_\mathrm{HK}}\rangle=-5.62$, it is suspected that Gliese 12 is too quiet to confidently ascertain the rotational period using RV data alone. GP-free, multi-Keplerian fits were also run as discussed in Section \ref{sec:secondplan}.

\section{Joint photometry and RV modelling}
\label{sec:joint}

We used \texttt{PyORBIT}\footnote{\url{https://github.com/LucaMalavolta/PyORBIT}} \citep{Malavolta2016,Malavolta2018} to perform a joint fit of the space-based transit observations, radial velocities, and the spectroscopic activity index H$\alpha$. We deliberately excluded the ground-based MuSCAT2 and MuSCAT3 photometry from the analysis, as incorporating them significantly increases the computational cost of the fit. More importantly, ground-based light curves are often affected by systematics, such as atmospheric effects, airmass trends, and instrumental noise. These systematic effects can introduce biases in the determination of transit depths, particularly for shallow transits like this one. In contrast, space-based observations offer higher photometric stability and precision, making them more reliable for this kind of analysis. Transit modelling was carried out using \texttt{PyTransit}\footnote{\url{https://github.com/hpparvi/PyTransit}} \citep{Parviainen2015}, and the red noise in each light curve was accounted for using Gaussian processes with a Mat\'ern 3/2 kernel.

The RV systematics were also modelled using GPs, with selected hyperparameters shared with those used for the H$\alpha$ spectroscopic activity index in order to account for stellar activity. We selected H$\alpha$ as the preferred activity indicator because its GLS periodogram \citep{Zechmeister2009} showed a prominent peak near the stellar rotation period derived from ground-based photometry. We adopted a quasi-periodic with cosine (QPC) kernel \citep{Perger2021}, where the characteristic period ($P_{\mathrm{rot}}$), the active-region decay timescale ($\lambda$), and the coherence scale ($w$) were modelled as common (and free) hyperparameters across both the RV and H$\alpha$ datasets. Although we possessed an independent estimate of the stellar rotation period based on long-term photometric monitoring, we imposed a uniform prior on $P_{\mathrm{rot}}$ in the range [60\,d, 200\,d] in order to assess whether the GP model could recover the photometric value without direct constraint and due to the seasonal variability found in the photometry.

To speed up the analysis, we limited the \tess data to 12-hour time windows centred on the predicted mid-transit times. For both the \tess and \cheops transit light curves, the following parameters were fitted: the central time of transit ($T_0$), orbital period ($P$), eccentricity ($e$), and argument of periastron ($\omega$), using the parameterisation from \cite{Eastman2013} ($\sqrt{e}\sin{\omega}$, $\sqrt{e}\cos{\omega}$). We also fitted the quadratic limb-darkening coefficients ($u_1$, $u_2$) independently for each instrument, constrained by normal priors based on predicted values from \texttt{LDTK}\footnote{\url{https://github.com/hpparvi/ldtk}} \citep{Husser2013,Parviainen2015b}, as well as the impact parameter ($b$) and the planet-to-star radius ratio ($R_\mathrm{p}/R_\star$). We also fitted the stellar density ($\rho_\star$) using a normal prior based on the stellar parameters presented by \citep{Kuzuhara2024}. Each light curve included a white noise term as an additional free parameter.

For the RV data, we included instrument-specific offsets and jitter terms as free parameters in the model. In addition to these offsets, we fitted for $P$, $e$ (using the parametrisation $\sqrt{e}\sin{\omega}$, $\sqrt{e}\cos{\omega}$), and $K$. As described above, several GP hyperparameters were shared between the RV and H$\alpha$ time series. Specifically, the kernel amplitude ($H$) and the amplitude of the cosine component ($C$) were fitted independently for each time series, while $P_{\mathrm{rot}}$, $\lambda$, and $w$ were treated as common parameters and fitted jointly.

The joint fit was performed in two stages. First, we applied global optimisation of the parameters using \texttt{PyDE}. Starting from the optimal solution, the parameter space was subsequently explored using \texttt{emcee}, running for $2.5 \times 10^6$ iterations to meet the convergence criteria of \texttt{PyORBIT}. The final fitted orbital and planetary parameters and corresponding priors are listed in Tables~\ref{tab:fit_results_planet} and \ref{tab:fit_priors_planet}, respectively. Similarly, the stellar and instrumental priors and resulting values are listed in Tables \ref{tab:fit_priors_starinst} and \ref{tab:fit_results_starinst}.

Our joint fit yields results for Gliese 12\,b that are consistent with those obtained from the photometry-only and informed RV analyses. The simultaneous modelling of the RVs and the H$\alpha$ activity index using the QPC kernel produced a rotation period of $P_\mathrm{rot} = 172.0_{-1.6}^{+1.5}\,\mathrm{d}$. This value is close to twice the rotation period inferred from long-term photometry ($P_\mathrm{rot} \sim 85\,\mathrm{d}$). We note that $P_\mathrm{rot}$ here refers to the characteristic period recovered by the GP model and is not necessarily the stellar rotation period itself. The QPC kernel, as described by \citet{Perger2021}, is designed to account for quasi-periodic signals at both $P_\mathrm{rot}$ and its first harmonic ($P_\mathrm{rot}/2$), which commonly arise due to spot configurations that produce symmetric signals that repeat every half rotation. In this context, the fitted period of $\sim$172\,d probably corresponds to the harmonic of a true rotation period near 85\,d. Tests using the QPC kernel on the RV and H$\alpha$ datasets separately confirm that the kernel can recover the 85-day period when a uniform prior in the range [60\,d, 100\,d] is adopted for $P_\mathrm{rot}$. A stellar rotation period significantly longer than 100\,d appears unlikely, as such long periods are rare among M dwarfs \citep[see][]{Newton2016b,Shan2024}. We therefore consider a rotation period close to 85\,d as the more physically plausible scenario for Gliese 12.

\section{Discussion}
\label{sec:discussion}

All the orbital and planetary parameters are consistent across the different fits. For the upcoming discussion of Gliese 12\,b, we adopted as final parameters the values from the joint fit. Thanks to our extensive RV campaign, we now have the first definitive mass for Gliese 12\,b. Having both a precise radius and mass allows for an in-depth study of a planet that is not possible without a mass measurement, see \citet{Batalha2019, Egger2024}. A bulk density can be obtained, which in turn can say something about the type of planet Gliese 12\,b is and how that compares with other currently known exoplanets. We can also go one step further and infer the possible interior composition of Gliese 12\,b as well as the possibility of it having an atmosphere. We investigate all of this in the upcoming sections. Finally, we examine the data for evidence of additional, detectable planets in the system.

\subsection{Gliese 12\,b as an exo-Venus}
\label{sec:venus}

\begin{figure}
    \centering
    \includegraphics[width=\linewidth]{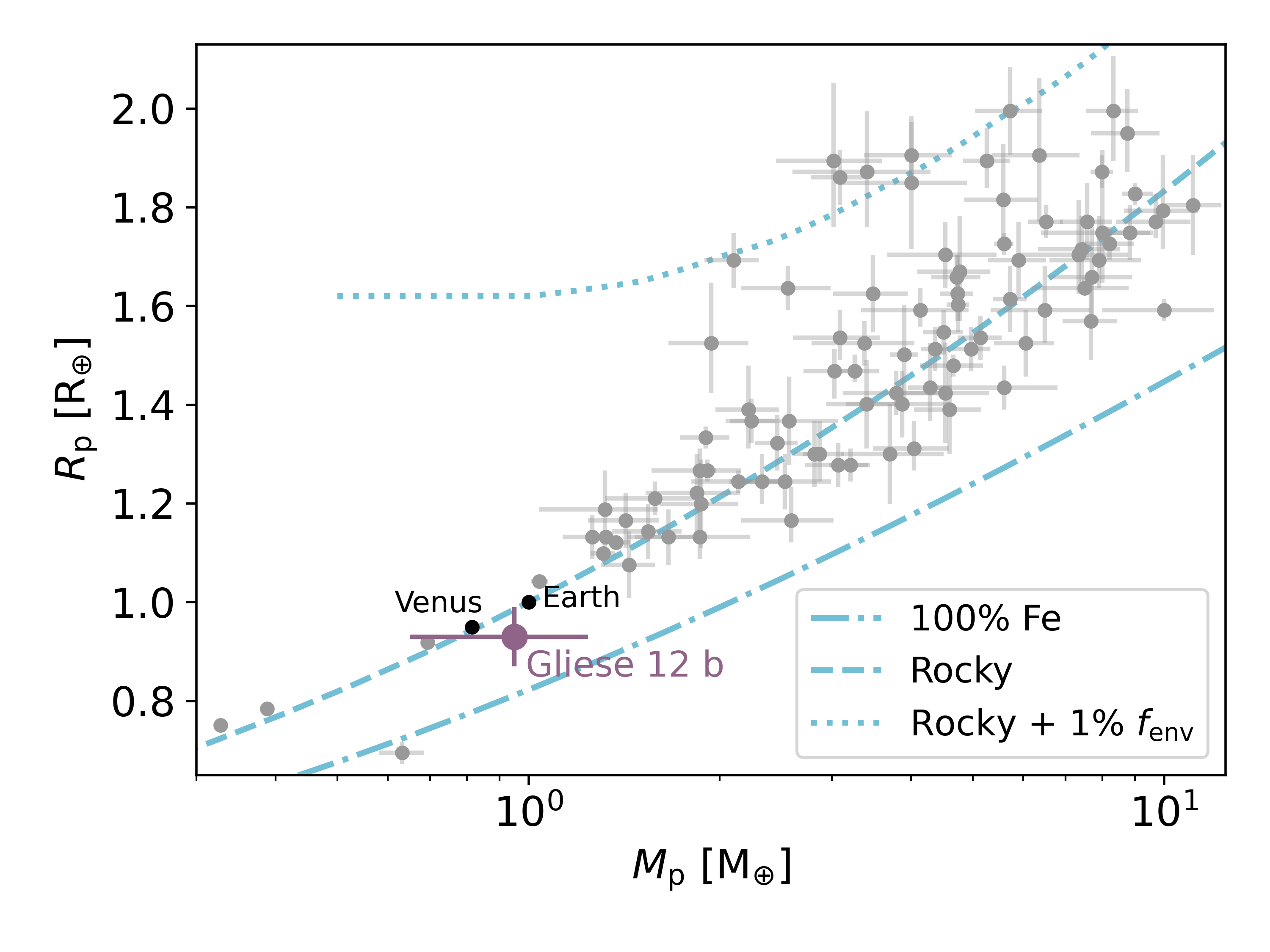}
    \caption{Mass-radius diagram of all small planets ($R_\mathrm{p}\leq2\,\mathrm{R_\oplus}$) in the PlanetS database (grey), Earth and Venus (black), and Gliese 12\,b (large purple). The 100 per cent Fe and rocky compositional lines are taken from \citet{Zeng2019}, and the rocky + 1 per cent $f_\mathrm{env}$ line is from \citet{Chen2016}.}
    \label{fig:mr}
\end{figure}

Gliese 12\,b occupies a unique region of parameter space due to its distinctive physical properties. For comparison reasons, we used the well-curated PlanetS database \citep{Parc2024} maintained on the DACE platform\footnote{\url{https://dace.unige.ch/exoplanets/}}. We further limited it to all planets with a radius less than $2$\,R$_\oplus$. Fig.~\ref{fig:mr} shows the mass--radius diagram of this sample of small planets. It is directly obvious that Gliese 12\,b lies in the same part of parameter space as Earth and Venus with a radius closer to Venus. Its density makes it marginally denser than our rocky Solar System planets.

The orbital period of Gliese 12\,b is much smaller than those of Earth and Venus. However, since the host star is an M dwarf, Gliese 12\,b is actually a temperate planet, too. We have calculated $T_\mathrm{eq}$ following \citet{Cowan2011}. As the albedo is unknown, we have to make an assumption. We calculated $T_\mathrm{eq}$ three times, using a Bond albedo of $0$, as well as both Earth's and Venus' albedo ($0.31$ and $0.77$ respectively). We found that the $T_\mathrm{eq}$ of Gliese 12\,b is between 317 and 219\,K. When using Venus' albedo, the $T_\mathrm{eq}$ of Gliese 12\,b is extremely close to that of Venus itself. Combined with the more Venus-like radius, we could conclude that this planet is possibly Venus-like.

\begin{figure}
    \centering
    \includegraphics[width=\linewidth]{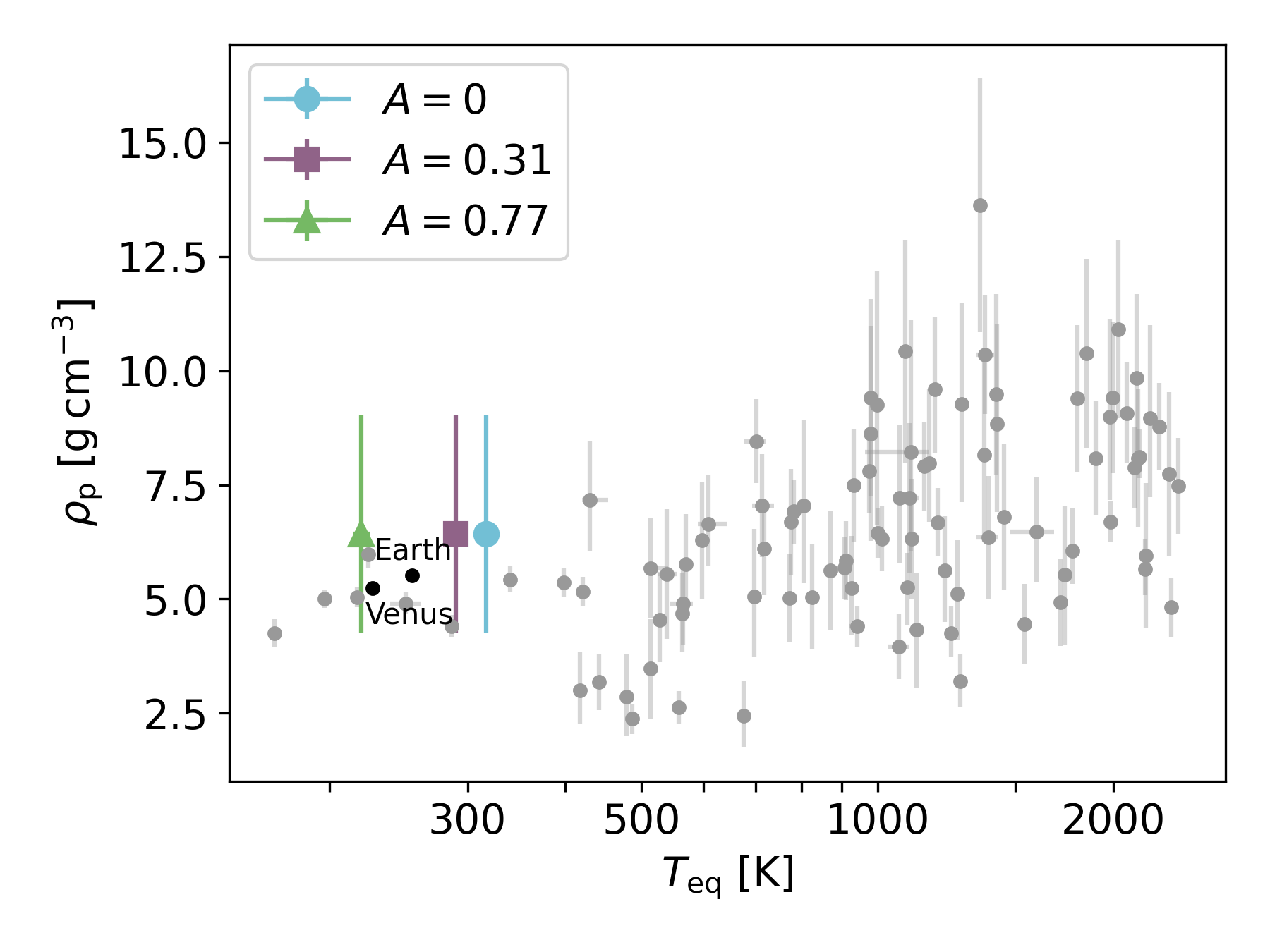}
    \caption{Planetary density versus equilibrium temperature for the same sample of planets as in Fig.~\ref{fig:mr}. For Gliese 12\,b, different Bond albedo values were used: $A=0$, Earth's albedo ($A=0.31$), and Venus' albedo ($A=0.77$). The planets in grey are those from the PlanetS database, with the TRAPPIST-1 planets lying in a similar region to Gliese 12\,b.}
    \label{fig:teq_dens}
\end{figure}

Fig.~\ref{fig:teq_dens} shows the different positions of Gliese 12\,b on a planet density versus equilibrium temperature diagram. We caveat that different albedos may have been used for the PlanetS database. In any case, whichever the true $T_\mathrm{eq}$, Gliese 12\,b occupies a sparsely populated area in this parameter space, making it a truly remarkable characterised exoplanet. The only well-characterised ($\sigma_{R_\mathrm{p}}\leq 8$ per cent; $\sigma_{M_\mathrm{p}}\leq 25$ per cent) planets that also occupy a similar region in parameter space are LHS 1140\,b \citep{Cadieux2024}, a super-Earth characterised by ESPRESSO, and the TRAPPIST-1 planets \citep{Gillon2017, Agol2021} where mass measurements come from transit timing variations.

As its host star is also one of the closest known planet host stars, there are plenty of opportunities for continued follow up of this exo-Venus. Additional information on atmospheric properties, and further and deeper characterisation of the system will reveal how similar (or discrepant) Gliese 12\,b is from Venus itself.

\subsection{Interior composition modelling}

Having a planetary mass and radius for Gliese 12\,b, we calculated a mean bulk density of 
$6.4 \pm 2.4\,\mathrm{g\,cm^{-3}}$, suggesting it is ever so slightly denser than Earth and Venus. We can also model the possible interior compositions for this small planet. For this purpose, we used the publicly available interior modelling code \texttt{plaNETic}\footnote{\url{https://github.com/joannegger/plaNETic}} \citep{Egger2024}. This framework utilises a neural network that was trained on the \texttt{BICEPS} forward model \citep{Haldemann2024}, combined with a full grid accept-reject sampling scheme. \texttt{plaNETic} has been shown to be fast as well as reliable when characterising the possible interior compositions of small planets.

Within \texttt{plaNETic}, a planet can have up to three layers: an inner core made of iron (Fe) and sulphur (S), a rocky mantle composed of oxidised silicon (Si), magnesium (Mg), and Fe, and a volatile layer uniformly mixed between water and pure hydrogen-helium. It thus does not have a separate water layer.

We assumed in the model that the planet formed outside the ice line since a H/He envelope as expected from a formation within the ice line would not be stable against evaporation as shown in Section~\ref{sec:atmo_loss} and all water was thus accreted in the form of ice \citep[see ][ for more detail]{Egger2024}. Although we cannot rule out the possibility of an in-situ formation, we model the interior of Gliese 12\,b based on the above assumption as a demonstration. Since Gliese 12\,b is either a bare core planet or has a high mean molecular weight envelope, we model this planet with the water-rich prior in \texttt{plaNETic}.

Different compositional priors can further be set based on the host star. There are preliminary studies showing possible links between host star and planet composition \citep[e.g. ][]{Thiabaud2015,Mortier2020,Adibekyan2021}. However, the exact link is still unknown, which is why \texttt{plaNETic} allows for the different priors. The relevant priors relate to the Si/Mg/Fe ratios. They can be sampled uniformly, have a prior where the planet ratios are assumed to be equal to the star's \citep{Thiabaud2015}, or have a prior assuming the planet is iron-enriched compared to the star  as described by equation S5 of \citet{Adibekyan2021}. Within \texttt{plaNETic}, abundances are recorded in the form 10$^{[\mathrm{X/Fe}]_\star}$ in the posteriors, allowing us to compare the abundance priors, which are $0.88\pm0.25$ and $0.30_{-0.11}^{+0.21}$ for X$=$Si, and $0.79\pm0.19$ and $0.27_{-0.09}^{+0.17}$ for X$=$Mg for the equal and iron-enriched prior respectively. We used the stellar abundances derived in Section~\ref{sec:star} as priors, where we used Ti as a reliable proxy for Si. The quality of silicon lines for stars with temperatures similar to that of Gliese 12 becomes degraded in comparison to the quality in hotter stars, in which case, the silicon value may be erroneous. The use of Ti as a proxy is justifiable due to it being an alpha element, as is Si; both elements form via the same process during stellar nucleosynthesis, and should therefore follow similar trends in abundances. We selected Ti in particular as there are many resolvable lines, amenable to analysis.

\begin{figure*}
    \centering
    \includegraphics[width=\linewidth]{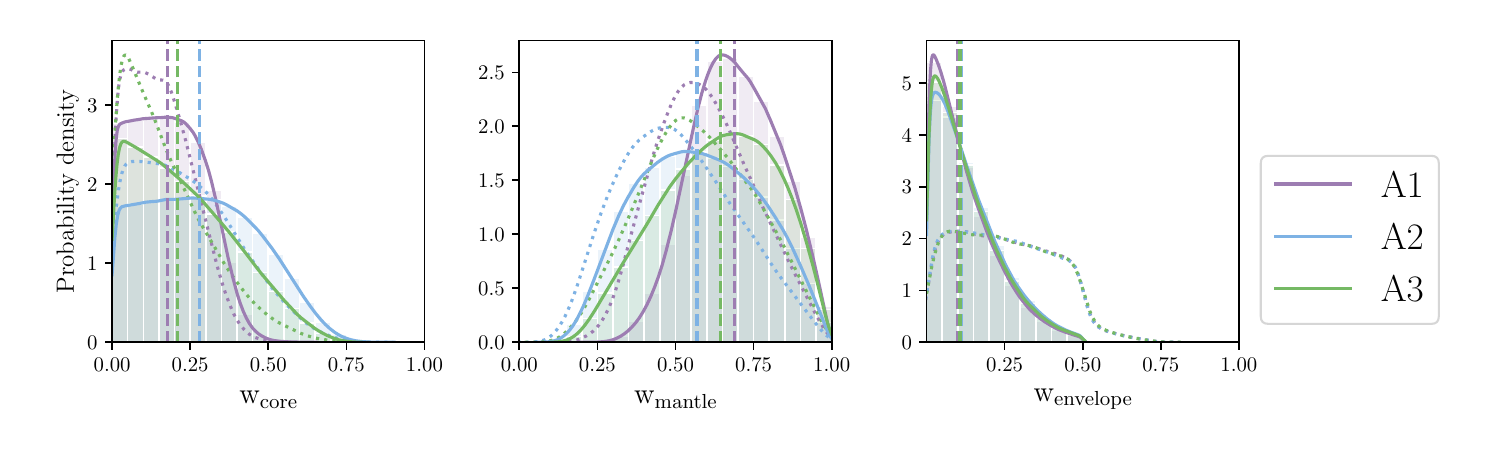}
    \caption{Posterior distributions (solid lines) of the interior structure of Gliese 12\,b, showing, from left to right, the mass fractions of the inner core, mantle, and volatile components. The different colours represent the three different priors for the planetary Si/Mg/Fe ratios: stellar (purple, A1), iron-enriched (blue, A2), and uniform (green, A3). Dotted lines indicate the priors and vertical dashed lines show the median of each posterior distribution.}
    \label{fig:planetic}
\end{figure*}

The resulting mass fractions of the different layers are shown in Fig.~\ref{fig:planetic}. As expected for this planet, the volatile mass fraction is small and consistent with 0. The core mass fraction is loosely constrained with median values between 20 and 30 per cent, depending on the prior. This is in line with the known core mass fraction of Venus (32 per cent).

\subsection{Atmospheric mass loss}
\label{sec:atmo_loss}

Gliese 12\,b is also an intriguing target for transmission spectroscopy in the Lyman-$\alpha$ line. Lyman-$\alpha$ observations can reveal escaping hydrogen gas during exoplanet transits \citep{VidalMadjar2003}, but interstellar absorption and geocoronal emission often obscure the line core, leaving observations to be sensitive only in the high-velocity line wings \citep[e.g.][]{Ehrenreich2015}. Nearby systems with high systemic RVs offer both higher flux and weaker interstellar absorption, and offset the interstellar absorption line core from the stellar emission line core. This allows to reveal information content at lower velocities and these targets are thus ideal for these studies.

At a distance of $12.2$\,pc and a systemic velocity of $51.2$\,km\,s$^{-1}$, the Gliese 12 system satisfies both of these criteria. Using the serendipitous detection of X-ray emission from Gliese 12 by \textit{XMM-Newton} \citep[$F_\mathrm{X} = 6.8\pm2.4\times10^{-15}$ erg~cm$^{-2}$~s$^{-1}$;][]{Webb2020} and the \citet{Linsky2013} relation between Lyman-$\alpha$ and X-ray flux, we expected Gliese 12 to have an intrinsic Lyman-$\alpha$ flux of $1.1\times10^{-13}$~erg~cm$^{-2}$~s$^{-1}$. Based on the modest column density of neutral hydrogen predicted by the Colorado LISM model \citep{Redfield2000} of $N_\mathrm{H} = 1.1\times10^{18}$~cm$^{-2}$ and the high systemic RV, we expected Gliese 12\,b to be a suitable target for Lyman-$\alpha$ transmission spectroscopy. It will soon be observed for this purpose by \textit{HST} GO 17600 (PI: Vissapragada). 

\begin{figure}
    \centering
    \includegraphics[width=\linewidth]{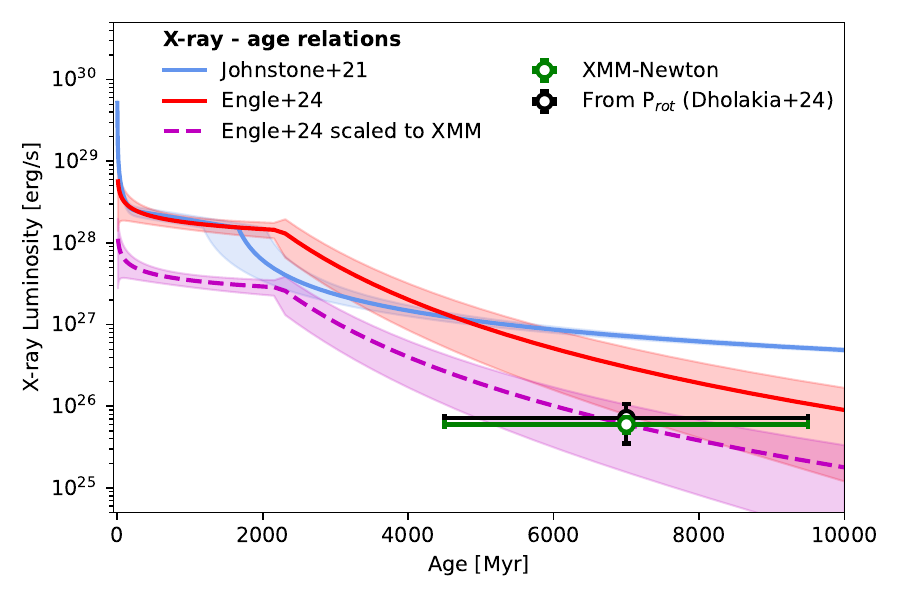}
    \caption{
        Comparison of several X-ray luminosity--age relations with the {\it XMM-Newton} observation of Gliese\,12 (green circle). The plotted X-ray age relations are: \citet[][blue line]{Johnstone21}, \citet[][red line]{Engle24}, and the same model scaled down to fit the {\it XMM-Newton} measurement from \citet{Kuzuhara2024}.
        The shaded regions correspond to the $1\sigma$ uncertainties in the models.
        The X-ray luminosity determined from the spin period measured by \citet{Dholakia2024} and the rotation-activity relation by \citet{Engle23} is also shown (black circle) and is in very good agreement with the measured X-ray emission.
    }
    \label{fig:star-evo}
\end{figure}

\begin{figure}
    \centering
    \includegraphics[width=\linewidth,trim={0 0 12.5cm 0},clip]{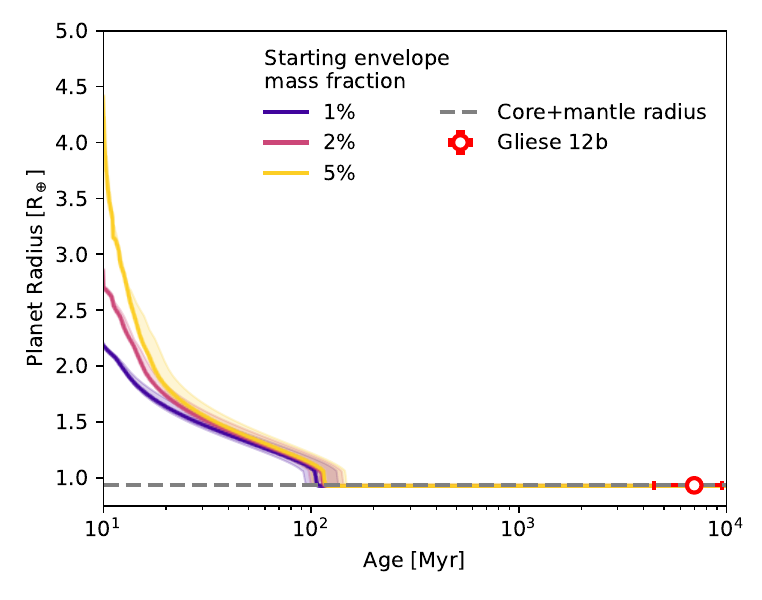}
    \includegraphics[width=\linewidth,trim={0 0 12.5cm 0},clip]{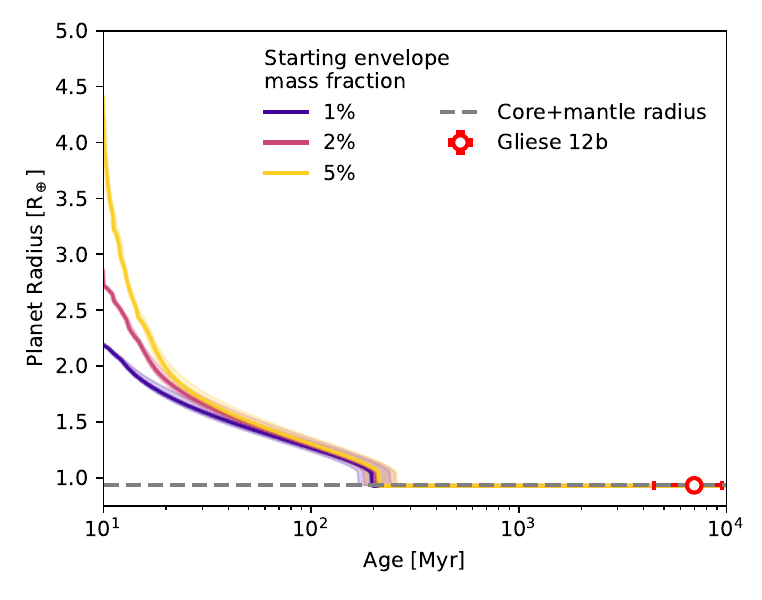}
    \caption{
        {\bf Top panel}: evolution of the total radius (core + mantle + envelope) of Gliese\,12\,b under atmospheric escape using the X-ray irradiation history predicted by \citet{Engle24}, with a range of initial atmospheres with mass fractions of 1 per cent (blue line), 2 per cent (red line), and 5 per cent (yellow line). 
        The shaded regions correspond to possible evaporation histories taking into account the $1\sigma$ uncertainties in the planet's radius, mass and X-ray irradiation history.
        The present-day total radius and age of Gliese\,12\,b is plotted as a red circle, and the size of its core and mantle as a dashed grey line.
        {\bf Bottom panel:} evolution of the total radius (core + mantle + envelope) of Gliese\,12\,b under atmospheric escape, following the top panel, using the X-ray irradiation history predicted by \citet{Engle24} scaled down by a factor of 5 to match the {\it XMM-Newton} observation.
    }
    \label{fig:planet-evo}
\end{figure}

We additionally simulated the past evaporation history of Gliese 12\,b in order to determine whether a H/He-rich envelope would have survived to the present day under X-ray irradiation from its host star.
To do so, we adopted the \texttt{photoevolver} code \citep{Fernandez23}, which combines three components to simulate the evaporation history of close-in exoplanets; these are (1) a model of the X-ray emission history of the host star, (2) an envelope structure model that links the mass fraction of the H/He atmosphere to its thickness, and (3) a model for the atmospheric escape rate.
Therefore, on each time-step, \texttt{photoevolver} computes the X-ray emission at the current simulation age using the X-ray emission model. Then, the corresponding incident X-ray flux of the planet is fed into the mass loss model to obtain a mass loss rate. The amount of mass lost during the time-step is then removed from the planet's gaseous atmosphere; and, finally, the envelope thickness is recalculated using the envelope structure model.

Regarding the X-ray emission history model, in Fig.~\ref{fig:star-evo} we compared the {\it XMM-Newton} detection of Gliese\,12 with the X-ray luminosity--age relations by \citet{Johnstone21}, which were based on a rotational evolution model fitted to open clusters, and by \citet{Engle24}, which were based on empirical fits to populations of M dwarfs.

At the age of Gliese\,12 (see Tab.\,\ref{tab:stellar_params}), the model by \citet{Engle24} predicted an X-ray luminosity of $3.0^{+10}_{-2.0}\times10^{26}$\,erg\,s$^{-1}$ (energy range\,0.1--2.5\,keV), whilst the model by \citet{Johnstone21} predicted $6.6^{+5.0}_{-2.0}\times10^{26}$\,erg\,s$^{-1}$ in the same energy range.
Additionally, \citet{Kuzuhara2024} presented {\it XMM-Newton} measurements of Gliese\,12 and reported an X-ray luminosity of $6.0\pm1.3\times10^{25}$\,erg\,s$^{-1}$ (range 0.124--2.5\,keV), which corresponds to $6.2\pm1.3\times10^{25}$\,erg\,s$^{-1}$ when extrapolated to the energy range of 0.1--2.5\,keV used by the models above (using the {\it WebPIMMS} tool\footnote{\url{https://heasarc.gsfc.nasa.gov/Tools/w3pimms_help.html}}).

Overall, we found the model by \citet{Engle24} more consistent with the {\it XMM-Newton} X-ray luminosity presented by \citet{Kuzuhara2024}, within a factor of 5, and thus we favoured it over the model by \citet{Johnstone21}, which is 11 times brighter.
We additionally considered an alternative X-ray emission history for Gliese\,12 where we scaled down the model by \citet{Engle24} to fit the {\it XMM-Newton} measurement presented by \citet{Kuzuhara2024}.

We also considered the contribution from the stellar extreme ultraviolet (EUV) emission, which can have a significant impact on atmospheric escape on longer timescales than X-ray alone \citep{OwenJackson2012:EUV, King2018:EUV, Sanz-Forcada2025}.
We adopted the X-ray--EUV relations by \citet{King2018:EUV} to compute the EUV emission history (13.6--100\,eV) from the X-ray emission history models (0.1--2.5\,keV) described above. In our simulations, we then considered the contribution from both the X-ray and EUV irradiation when computing the mass loss history of the planet.

Regarding the envelope structure model, we adopted the models presented by \citet{Chen2016}. They adapted the {\tt MESA} code \citep{Paxton2011:MESA} to build mass-radius relations for gaseous atmospheres on planets from Earth-mass to Jupiter-mass. In this work, we adopted the polynomial fit that they performed to their simulation results \citep[][Eqn. 5]{Chen2016}, valid for envelope mass fractions of 0.01--20 per cent and planet masses of 1--20\,M$_\oplus$.

Finally, regarding the mass loss model, we adopted the models presented by \citet{Kubyshkina18}, which accounts for both core-powered mass loss as well as energy-limited and recombination-limited escape from both X-ray and EUV photons, and is based on hydrodynamic simulations.
In this work, we adopted the pre-computed grid of planet parameters and interpolation routine presented by \citet{KubyshkinaFossati21}, valid for planet masses of 1--39\,M$_\oplus$ and planet radii of 1--10\,R$_\oplus$.

Motivated by its high density, we assumed that Gliese 12\,b has no significant gaseous envelope at present. We then considered a number of initial envelopes on the planet, with mass fractions of 1 per cent, 2 per cent, and 5 per cent. We did not consider larger initial atmospheres as these would have been stripped off and truncated by boil-off processes shortly after dispersal of the protoplanetary disc \citep{Rogers24}.

We thus simulated the past evaporation history of Gliese 12\,b in each scenario. Our photoevolver calculations adopted the 4th-order Runge-Kutta integration method with a maximum time step of 1\,Myr, in the interval between 10\,Myr and 1\,Gyr.
Additionally, we considered the gaseous envelope to be fully stripped when it reaches a mass fraction below 0.01 per cent, which corresponds to the minimum value the envelope structure model by \citet{Chen2016} is rated for.

In Fig.~\ref{fig:planet-evo} we show the results from our simulations, from 10\,Myr to 10\,Gyr. On the top panel we show atmospheric evolution under the X-ray emission history predicted by \citet{Engle24} (and its corresponding EUV emission history from the relations by \citet{King2018:EUV}), and on the bottom panel, the atmospheric evolution under the same emission history, but scaled down by a factor of 5 to match the {\it XMM-Newton} observation of Gliese\,12 as we do not know the true X-ray history of the star. We found that all H/He envelopes are fully stripped from Gliese 12\,b within 150 Myr, with none surviving to the present day. Even with the faintest X-irradiation model, the envelope lifetime is only extended to the age of 200\,Myr.

In our simulations, we neglected the contribution to atmospheric escape from additional mechanisms, such as stellar flares, coronal mass ejections, and magnetic interactions \citep[e.g.][]{Vidotto2013:star-planet-magnetism, Atri2021:flares-mass-loss, Hazra2025:CME-mass-loss}. These mechanisms would only strengthen mass loss rates, causing the planet's envelope to be lost even earlier in its life.

\subsection{Prospects for atmospheric characterization}

\begin{figure}
    \centering
    \includegraphics[width=0.49\textwidth]{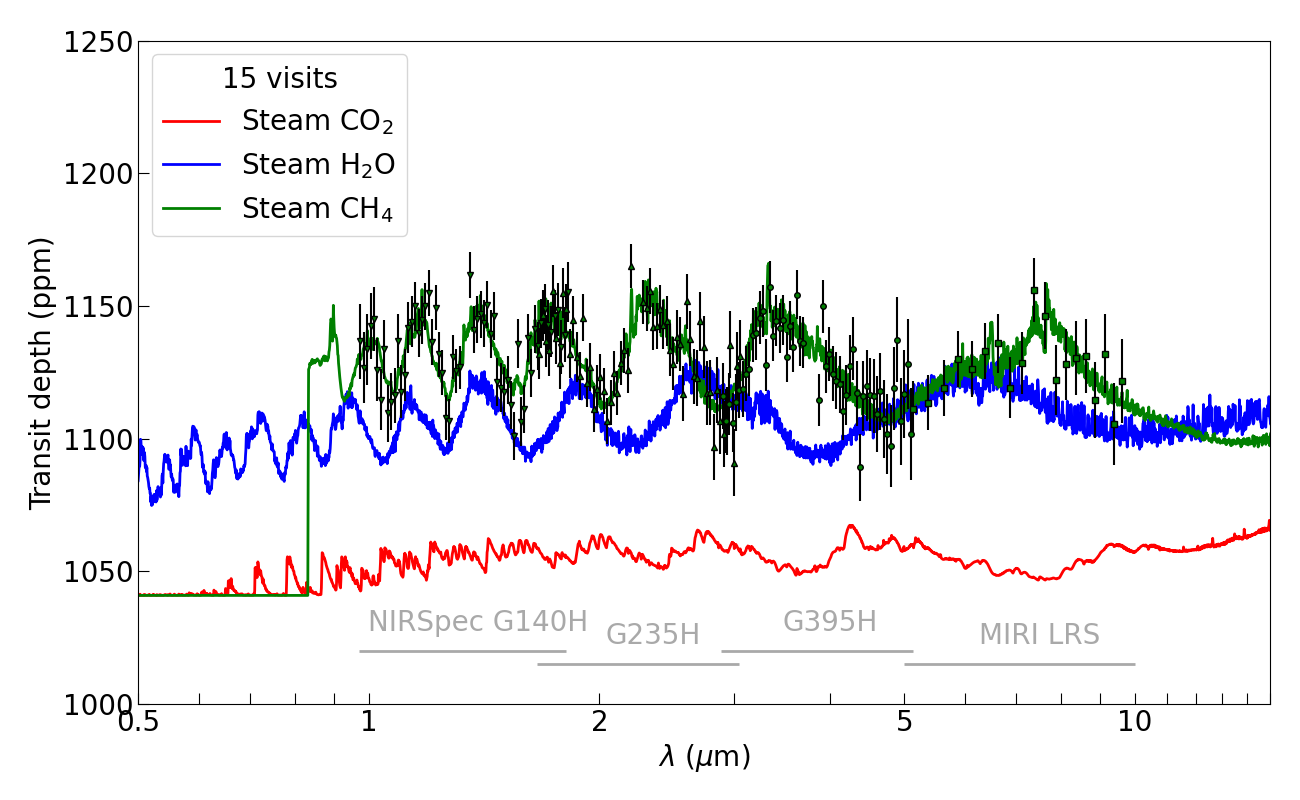}
    \includegraphics[width=0.49\textwidth]{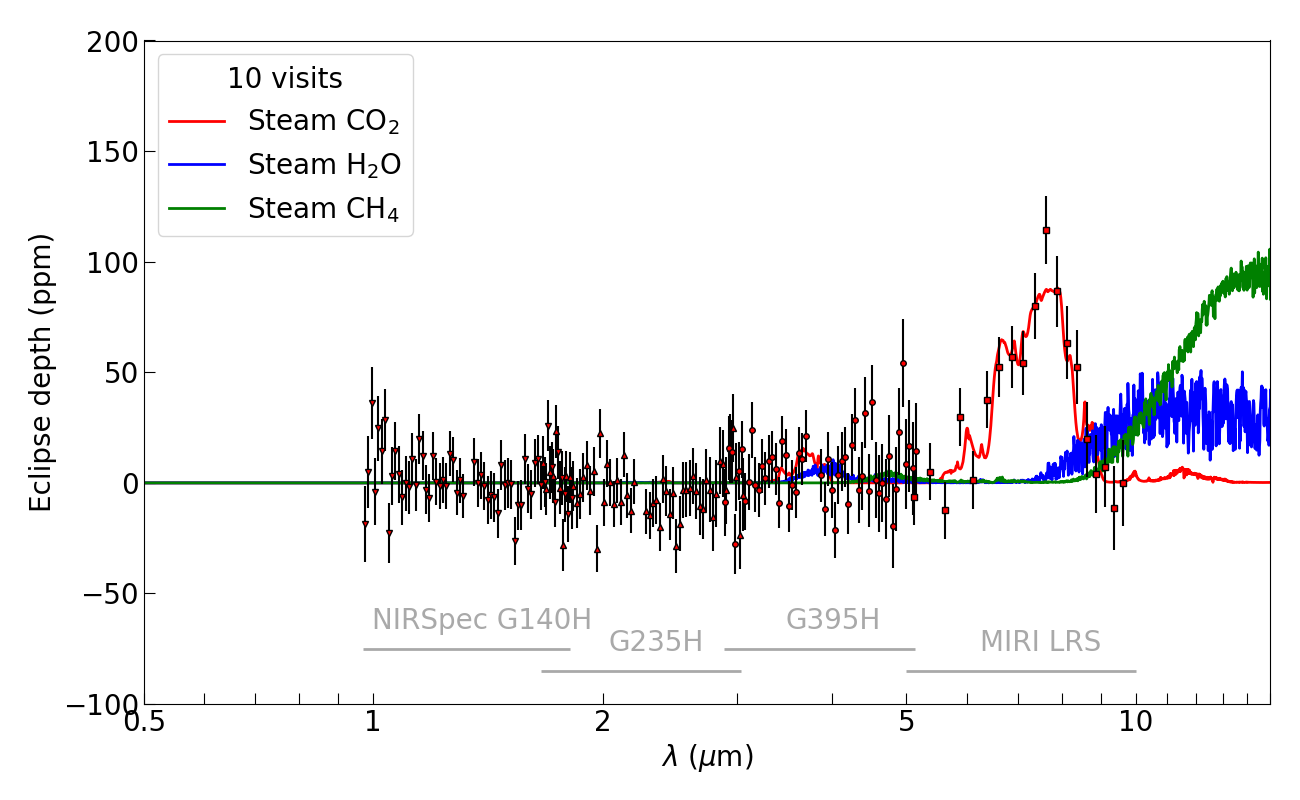}
    \caption{
        {\bf Top panel}: Modelled transmission spectra of Gliese 12\,b, assuming an atmospheric composition of 100 per cent CO$_2$ (red), 100 per cent H$_2$O (blue), and 100 per cent CH$_4$ (green). Points with error-bars show the simulated spectrum for the CH$_4$ composition, combining 15 transits with each of the NIRSpec G140H, G235H, G395H and MIRI LRS Slitless modes. 
        {\bf Bottom panel:} Analogous emission spectra. Points with error-bars show the simulated spectrum for the CO$_2$ composition, combining 10 eclipses with each of the NIRSpec G140H, G235H, G395H and MIRI LRS Slitless modes.
    }
    \label{fig:jwst_atmo}
\end{figure}

Given its high mean density and the intense XUV irradiation from its host star, Gliese 12\,b is unlikely to have retained a primordial or secondary atmosphere rich in H$_2$/He (see Section \ref{sec:atmo_loss}). Nonetheless, owing to the brightness of its host star, Gliese 12\,b stands as one of the top rocky targets for atmospheric follow-up with JWST. Following \citet{Kempton2018}, we used the planetary mass and radius, stellar radius, equilibrium temperature, and the apparent stellar magnitude in the $J$ band to calculate a Transmission Spectroscopy Metric (TSM) of 14$_{-4}^{+7}$, similar to that of TRAPPIST-1\,e, f, g, and h.

We conducted simulations of transmission and emission spectra for realistic atmospheric scenarios, as they would be observed with various JWST instrument configurations (Fig.~\ref{fig:jwst_atmo}). In particular, we modelled a Venus-like atmosphere consisting of 100 per cent CO$_2$, motivated by the planet's physical parameters resembling those of Venus (see Section \ref{sec:venus}); a steam H$_2$O envelope representing a water-world scenario; and a lighter steam CH$_4$ envelope featuring a reduced environment. We used \texttt{TauREx 3} for radiative transfer modelling \citep{Waldmann2015a,Waldmann2015b,Al-Refaie2021}, employing a temperature-pressure profile similar to that of Venus for the CO$_2$ atmosphere and profiles adapted from \citet{Kempton2023} for the H$_2$O and CH$_4$ compositions.
The absorption cross sections for H$_2$O, CO$_2$, and CH$_4$ were obtained from the ExoMol database \citep{Chubb2021}, which utilised line lists by \citet{Polyansky2018}, \citet{Rothman2010}, and \citet{Yurchenko2017}, respectively.
The corresponding JWST spectra were generated using \texttt{ExoTETHyS} \citep{Morello2021} for the NIRSpec G140H (0.97--1.83\,$\mu$m), G235H (1.66--3.07\,$\mu$m), and G395H (2.88--5.18\,$\mu$m) modes, as well as the MIRI LRS Slitless (5--10\,$\mu$m) mode, based on updated instrument response files. We included an additional 30 per cent noise contribution in the final error bars. The popular NIRISS SOSS mode was excluded due to detector saturation after the first group. We adopted spectroscopic binning with a resolving power of $\mathcal{R} \sim 100$ for the NIRSpec modes, following the recommendations of \citet{Carter2024} and \citet{Davey2025}, and a constant bin size of 0.25\,$\mu$m for the MIRI LRS, as suggested by \citet{Powell2024}. The predicted error bars for a single transit observation are 32–54 ppm (mean error 37 ppm) for NIRSpec G140H, 32–52 ppm (mean error 38 ppm) for NIRSpec G235H, 36–68 ppm (mean error 46 ppm) for NIRSpec G395H, and 40–61 ppm (mean error 48 ppm) for MIRI LRS. These uncertainties are comparable to or smaller than the expected spectral features in transmission, which reach $\lesssim$50 ppm in the most favourable CH$_4$-dominated scenario. The emission spectra are generally flat across the wavelength range covered by our instrumental simulations, except for a $\sim$90 ppm CO$_2$ feature in the Venus-like scenario.

Another possibility is to measure planetary emission at longer wavelengths. In fact, Gliese 12\,b is a target under consideration for the Rocky World DDT program \citep{Redfield2024}, potentially aiming to observe multiple secondary eclipses using a MIRI filter centred at 15\,$\mu$m. In all cases, multiple visits are required to robustly detect atmospheric features, with the optimal strategy, transmission or emission spectroscopy, depending on the actual atmospheric composition of Gliese 12\,b.

\subsection{Possibility of extra planets in the system}
\label{sec:secondplan}

The possibility of there being more planets within the system still remains, and further investigation is needed to reveal any previously undetected signals. We have performed additional RV fits using two eccentric Keplerian models to explore any potential multi-planet configurations of small planets with periods below 50\,days (based of the periodogram information). While a solution was found with a second Keplerian close to 34\,days, the parameters were poorly constrained and the single-Keplerian model with a GP applied to the RVs was preferred using the BIC. We obtained $\mathrm{BIC} = -1774.7$ versus $\mathrm{BIC} = -1846.5$ for the two-Keplerian model and the adopted model, respectively.

Ultimately, more data are required to constrain any existing secondary signals, and authenticate the system architecture. More RV data will further allow for the resolution of current degeneracies that exist between eccentricity, stellar activity, and any undetected planetary companions and make the mass measurement of Gliese 12\,b more robust. Obtaining more RVs with the precision of those within this study could reveal planets hidden below the current precision floor of the dataset.

\section{Conclusions}
\label{sec:conclude}


In this paper, we present new analyses of the transiting exo-Venus, Gliese 12\,b. We performed a photometry-only fit with \texttt{juliet}, an informed RV fit with \texttt{pyaneti}, and a joint fit using both datasets simultaneously with \texttt{PyOrbit}. All orbital and planetary parameters are consistent across fits and modelling software. 

With new \tess and \cheops observations, we report refined orbital period ($12.76\,\mathrm{d}$) and planetary radius measurements ($0.93 \pm 0.06 \,\mathrm{R_\oplus}$). Our updated radius value is consistent with the values reported in previous studies \citep{Dholakia2024, Kuzuhara2024}.

Using nearly 200 high-precision RV observations from HARPS-N, ESPRESSO, and CARMENES, we report the first definitive mass measurement of Gliese 12\,b. Owing to the extremely small RV semi-amplitude of Gliese 12\,b, we adopted a variety of modelling approaches dealing with the stellar signals in different ways. We detect an RV semi-amplitude of $0.67\pm0.21\,\mathrm{m\,s^{-1}}$ ($3.2\sigma$), yielding a mass measurement of $0.95_{-0.30}^{+0.29}\,\mathrm{M_\oplus}$. This measurement is among the smallest semi-amplitudes ever measured for a transiting planet and is pushing the boundaries of our current capabilities.

A thorough investigation was conducted into the activity signal of Gliese 12. Analysis of long-term photometric monitoring from ASAS-SN, LCOGT, TJO, and E-EYE -- as well as a DCF analysis of the RV and H$\alpha$ data -- provided further evidence towards a stellar rotation period of approximately 85\,d (see Fig.~\ref{fig:longterm-photometry}). This work highlights the importance of the rigorous treatment of stellar variability when searching for the smallest exoplanets in RV data.

Gliese 12\,b stands out with its physical and orbital properties. We investigated the nature of Gliese 12\,b, by studying its interior structure, atmospheric mass loss history and future atmospheric prospects. Interior structure modelling revealed that Gliese 12\,b is likely to be similar to Venus in composition (predominantly rocky make-up). The bulk properties of this planet place it in a unique region of parameter space, distinguishing it from other exoplanets with similar properties (see Figs.~\ref{fig:mr} and \ref{fig:teq_dens}). Furthermore, we demonstrated that the planet is unlikely to have retained a primordial gaseous envelope, due to the X-ray emission history of its host star. Given its intriguing properties, Gliese 12\,b presents a promising opportunity for Lyman-$\alpha$ spectroscopic follow-up studies, and is already being observed with \textit{HST} for this reason. It is also a fantastic target for the Rocky Worlds DDT programme on \textit{JWST}.

\section*{Acknowledgements}

We thank the referee for a constructive and timely report that improved the quality of this manuscript.

This work is based on observations made with the Italian Telescopio Nazionale Galileo (TNG) operated on the island of La Palma by the Fundaci\'on Galileo Galilei of the INAF (Istituto Nazionale di Astrofisica) at the Spanish Observatorio del Roque de los Muchachos of the Instituto de Astrofisica de Canarias. The HARPS-N project was funded by the Prodex Program of the Swiss Space Office (SSO), the Harvard University Origin of Life Initiative (HUOLI), the Scottish Universities Physics Alliance (SUPA), the University of Geneva, the Smithsonian Astrophysical Observatory (SAO), the Italian National Astrophysical Institute (INAF), University of St. Andrews, Queen’s University Belfast, and University of Edinburgh.

CARMENES is an instrument at the Centro Astron\'omico Hispano en Andaluc\'ia (CAHA) at Calar Alto (Almer\'{\i}a, Spain), operated jointly by the Junta de Andaluc\'ia and the Instituto de Astrof\'isica de Andaluc\'ia (CSIC). CARMENES was funded by the Max-Planck-Gesellschaft (MPG), the Consejo Superior de Investigaciones Cient\'{\i}ficas (CSIC), the Ministerio de Econom\'ia y Competitividad (MINECO) and the European Regional Development Fund (ERDF) through projects FICTS-2011-02, ICTS-2017-07-CAHA-4, and CAHA16-CE-3978, and the members of the CARMENES Consortium (Max-Planck-Institut f\"ur Astronomie, Instituto de Astrof\'{\i}sica de Andaluc\'{\i}a, Landessternwarte K\"onigstuhl, Institut de Ci\`encies de l'Espai, Institut f\"ur Astrophysik G\"ottingen, Universidad Complutense de Madrid, Th\"uringer Landessternwarte Tautenburg, Instituto de Astrof\'{\i}sica de Canarias, Hamburger Sternwarte, Centro de Astrobiolog\'{\i}a and Centro Astron\'omico Hispano-Alem\'an),  with additional contributions by the MINECO, the Deutsche Forschungsgemeinschaft (DFG) through the Major Research Instrumentation Programme and Research Unit FOR2544 ``Blue Planets around Red Stars'', the Klaus Tschira Stiftung, the states of Baden-W\"urttemberg and Niedersachsen, and by the Junta de Andaluc\'{\i}a.

This work is based on observations collected with ESPRESSO at the European Southern Observatory under ESO programme 113.26RH. 

{\it CHEOPS} is an ESA mission in partnership with Switzerland with important contributions to the payload and the ground segment from Austria, Belgium, France, Germany, Hungary, Italy, Portugal, Spain, Sweden, and the United Kingdom. The {\it CHEOPS} Consortium would like to gratefully acknowledge the support received by all the agencies, offices, universities, and industries involved. Their flexibility and willingness to explore new approaches were essential to the success of this mission. {\it CHEOPS} data analysed in this article will be made available in the {\it CHEOPS} mission archive (\url{https://cheops.unige.ch/archive_browser/}).

This paper made use of data collected by the \textit{TESS} mission and are publicly available from the Mikulski Archive for Space Telescopes (MAST) operated by the Space Telescope Science Institute (STScI). Funding for the \textit{TESS} mission is provided by NASA’s Science Mission Directorate. We acknowledge the use of public \textit{TESS} data from pipelines at the \textit{TESS} Science Office and at the \textit{TESS} Science Processing Operations Center. Resources supporting this work were provided by the NASA High-End Computing (HEC) Program through the NASA Advanced Supercomputing (NAS) Division at Ames Research Center for the production of the SPOC data products. This research has made use of the Exoplanet Follow-up Observation Programme (ExoFOP; DOI: 10.26134/ExoFOP5) website, which is operated by the California Institute of Technology, under contract with the National Aeronautics and Space Administration under the Exoplanet Exploration Programme.

This article is based on observations made with the MuSCAT2 instrument, developed by ABC, at Telescopio Carlos Sánchez, operated on the island of Tenerife by the IAC in the Spanish Observatorio del Teide.

This paper is based on observations made with the MuSCAT3 instrument, developed by the Astrobiology Center and under financial supports by JSPS KAKENHI (JP18H05439) and JST PRESTO (JPMJPR1775), at Faulkes Telescope North on Maui, HI, operated by the Las Cumbres Observatory.

This work has made use of data from the European Space Agency (ESA) mission {\it Gaia} (\url{https://www.cosmos.esa.int/gaia}), processed by the {\it Gaia} Data Processing and Analysis Consortium (DPAC, \url{https://www.cosmos.esa.int/web/gaia/dpac/consortium}). Funding for the DPAC has been provided by national institutions, in particular the institutions participating in the {\it Gaia} Multilateral Agreement.

This publication makes use of The Data \& Analysis Center for Exoplanets (DACE), which is a facility based at the University of Geneva (CH) dedicated to extrasolar planets data visualisation, exchange and analysis. DACE is a platform of the Swiss National Centre of Competence in Research (NCCR) PlanetS, federating the Swiss expertise in Exoplanet research. The DACE platform is available at \url{https://dace.unige.ch}.

This research has made use of the NASA Exoplanet Archive, which is operated by the California Institute of Technology, under contract with the National Aeronautics and Space Administration under the Exoplanet Exploration Program.

DAT acknowledges the support of the Science and Technology Facilities Council (STFC).

YNEE acknowledges support from a Science and Technology Facilities Council (STFC) studentship, grant number ST/Y509693/1.

AM, AAJ, BSL acknowledge funding from a UKRI Future Leader Fellowship, grant number MR/X033244/1. AM acknowledges funding from a UK Science and Technology Facilities Council (STFC) small grant ST/Y002334/1. 

TGW acknowledges support from the University of Warwick and UKSA.

GM acknowledges support from the Ram\'on y Cajal grant RYC2022-037854-I funded by MCIN/AEI/1144 10.13039/501100011033 and FSE+.

We acknowledge financial support from the Agencia Estatal de Investigaci\'on of the Ministerio de Ciencia e Innovaci\'on MCIN/AEI/10.13039/501100011033 and the ERDF “A way of making Europe” through projects PID2021-125627OB-C3[1:2], PID2023-152906NA-I00, and PID2022-137241NBC4[1:4], PID2019-107061GB-C61, CNS2023-144309 and PID2020-112949GB-I00 and from the Centre of Excellence “Severo Ochoa” award to the Instituto de Astrofisica de Canarias (CEX2021-001131-S).

This research was partially funded by The Israel Science Foundation through grant No. 1404/22.

This work was funded by the European Union (ERC, FIERCE, 101052347). Views and opinions expressed are however those of the author(s) only and do not necessarily reflect those of the European Union or the European Research Council. Neither the European Union nor the granting authority can be held responsible for them. This work was also supported by FCT - Fundação para a Ciência e a Tecnologia through national funds by these grants: UIDB/04434/2020 DOI: 10.54499/UIDB/04434/2020, UIDP/04434/2020 DOI: 10.54499/UIDP/04434/2020, PTDC/FIS-AST/4862/2020, UID/04434/2025. CEECIND/CP2839/CT0004, 2023.08117.CEECIND/CP2839/CT0004.

SD and APH acknowledge support from the DFG under Research Unit FOR2544 ``Blue Planets around Red Stars'' through projects DR 281/39-1 and HA 3279/14-1, respectively.

LN and LM acknowledge financial contribution from the INAF Large Grant 2023 ``EXODEMO''.

AVF acknowledges the support of the Institude of Physics through the Bell Burnell Graduate Scholarship Fund.

LM acknowledges the financial contribution from PRIN MUR 2022 project 2022J4H55R.

ASB acknowledges financial contribution from the European Union - Next Generation EU RRF M4C2 1.1 PRIN MUR 2022 project 2022CERJ49 (ESPLORA).

TT acknowledges support from the BNSF program ``VIHREN-2021'' project No. KP-06-DV/5.

IC acknowledges financial contribution from the European Union - Next Generation EU PNRR M4C2 1.2 project SOE2024.

MS thanks the Belgian Federal Science Policy Office (BELSPO) for the provision of financial support in the framework of the PRODEX Programme of the European Space Agency (ESA) under contract number C4000140754.

This work has been carried out within the framework of the NCCR PlanetS supported by the Swiss National Science Foundation under grants 51NF40\_182901 and 51NF40\_205606.

ACC acknowledges support from STFC consolidated grant number ST/V000861/1 and UKRI/ERC Synergy Grant EP/Z000181/1 (REVEAL).

EN acknowledges the support from the Deutsches Zentrum für Luft- und Raumfahrt (DLR, German Aerospace Center) - project number 50OP2502.

\section*{Data Availability}

The photometric data are publicly available via the respective archives. The RVs will be released via Vizier CDS. Further corner plots and posterior distributions are available upon request.



\bibliographystyle{mnras}
\bibliography{export-bibtex}

\vspace{1cm}
\noindent $^{1}$School of Physics and Astronomy, University of Birmingham, Edgbaston, Birmingham B15 2TT, UK\\
$^{2}$Department of Physics, University of Warwick, Gibbet Hill Road, Coventry, CV4 7LS, UK\\
$^{3}$Instituto de Astrofísica de Canarias (IAC), 38205 La Laguna, Tenerife, Spain\\
$^{4}$Departamento de Astrofísica, Universidad de La Laguna (ULL), 38206 La Laguna, Tenerife, Spain\\
$^{5}$Department of Physics \& Astronomy, McMaster University, 1280 Main St W, Hamilton, ON, L8S 4L8, Canada\\
$^{6}$Instituto de Astrof\'{i}sica de Andaluc\'{i}a (IAA), 18080, Granada, Spain\\
$^{7}$INAF -- Osservatorio Astronomico di Palermo, Piazza del Parlamento, 1, 90134 Palermo, Italy \\
$^{8}$Institut de Ciencies de l’Espai (ICE, CSIC), Campus UAB, Can Magrans s/n, 08193 Bellaterra, Spain\\
$^{9}$Institut d'Estudis Espacials de Catalunya (IEEC), 08860 Castelldefels (Barcelona), Spain\\
$^{10}$Space Research and Planetary Sciences, Physics Institute, University of Bern, Gesellschaftsstrasse 6, 3012 Bern, Switzerland \\
$^{11}$Carnegie Science Observatories, 813 Santa Barbara Street, Pasadena, CA 91101, USA\\
$^{12}$Centro de Astrobiología, CSIC-INTA, Camino Bajo del Castillo s/n, Villanueva de la Cañada, Madrid\\
$^{13}$Institut f\"ur Astrophysik und Geophysik, Georg-August-Universit\"at, Friedrich-Hund-Platz 1, 37077 G\"ottingen, Germany\\
$^{14}$Thueringer Landessternwarte Tautenburg, Germany \\
$^{15}$INAF -- Osservatorio Astrofisico di Torino, Via Osservatorio 20, 10025 Pino Torinese, Italy\\
$^{16}$Vereniging Voor Sterrenkunde (VVS), Oostmeers 122 C, 8000 Brugge, Belgium\\
$^{17}$Public observatory ASTROLAB IRIS, Provinciaal Domein “De Palingbeek”, Verbrandemolenstraat 5, 8902 Zillebeke, Ieper, Belgium\\
$^{18}$Centre for Astrophysics, University of Southern Queensland, Toowoomba, QLD 4350, Australia\\
$^{19}$DTU Space, Technical University of Denmark, Elektrovej 328, DK-2800 Kgs. Lyngby, Denmark\\
$^{20}$SUPA School of Physics and Astronomy, University of St Andrews, North Haugh, St Andrews, KY16 9SS, UK \\
$^{21}$Tata Institute of Funadamental Research, Mumbai, India\\
$^{22}$Department of Physics, Ariel University, Ariel 40700, Israel\\
$^{23}$Observatoire Astronomique de l'Universit\'e de Gen\`eve, Chemin Pegasi 51, CH-1290 Versoix, Switzerland\\
$^{24}$Fundaci\'on Galileo Galilei - INAF, Rambla Jos\'e Ana Fernandez P\'erez 7, E-38712 Bre\~na Baja, Tenerife, Spain\\
$^{25}$Space Telescope Science Institute, 3700 San Martin Drive, Baltimore, MD 20218, USA
$^{26}$Dipartimento di Fisica e Astronomia "Galileo Galilei", Università di Padova, Vicolo dell'Osservatorio 3, 35122, Padova, Italy\\
$^{27}$Istituto Nazionale di Astrofisica – Osservatorio Astronomico di Padova, Vicolo dell'Osservatorio 5, 35122, Padova, Italy\\
$^{28}$Department of Physics, University of Rome ``Tor Vergata'', Via della Ricerca Scientifica 1 I-00133 Roma, Italy\\
$^{29}$Max Planck Institute for Astronomy, K\"{o}nigstuhl 17, 69117 Heidelberg, Germany \\
$^{30}$Departamento de F\'{i}sica de la Tierra y Astrof\'{i}sica and IPARCOS-UCM (Instituto de F\'{i}sica de Part\'{i}culas y del Cosmos de la UCM), Facultad de Ciencias F\'{i}sicas, Universidad Complutense de Madrid, E-28040, Madrid, Spain\\
$^{31}$Department of Astronomy, University of Texas at Austin, 2515 Speedway, Austin, TX 78712, USA\\
$^{32}$SUPA, Institute for Astronomy, University of Edinburgh, Blackford Hill, Edinburgh, EH9 3HJ, UK\\
$^{33}$Centre for Exoplanet Science, University of Edinburgh, Edinburgh, EH9 3HJ, UK\\
$^{34}$Landessternwarte, Zentrum f\"ur Astronomie der Universit\"at Heidelberg, K\"onigstuhl 12, 69117 Heidelberg, Germany\\
$^{35}$Instituto de Astrof\'{\i}sica e Ci\^encias do Espa\c co, CAUP, Universidade do Porto, Rua das Estrelas, 4150-762 Porto, Portugal\\
$^{36}$Departamento de F\'{\i}sica e Astronomia, Faculdade de Ci\^encias, Universidade do Porto, Rua do Campo Alegre, 4169-007 Porto, Portugal \\
$^{37}$Space sciences, Technologies and Astrophysics Research (STAR) Institute, Université de Liège, Allée du 6 Août 19C, 4000 Liège, Belgium \\
$^{38}$Astrobiology Research Unit, Université de Liège, Allée du 6 Août 19C, B-4000 Liège, Belgium \\
$^{39}$Astrophysics, Geophysics, And Space Science Research Center, Ariel University, Ariel 40700, Israel\\
$^{40}$Department of Astronomy, Faculty of Physics, Sofia University ``St Kliment Ohridski'', 5 James Bourchier Blvd, 1164 Sofia, Bulgaria\\


\appendix

\section{Long-term photometry}

\begin{figure*}
    \centering
    \includegraphics[width=0.99\linewidth]{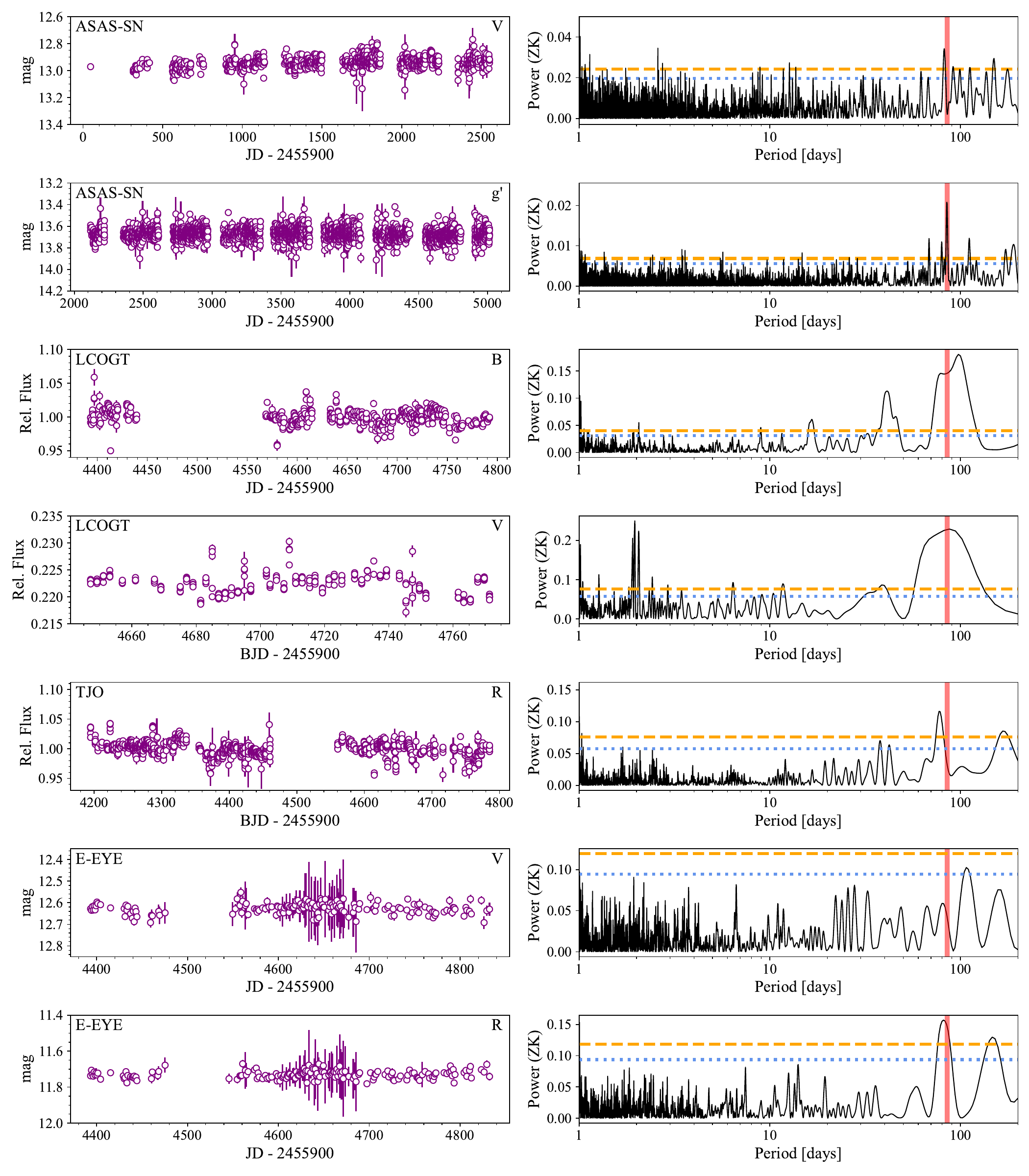}
    \caption{Gliese 12 long-term ground-based photometry and GLS periodograms of the time series. The vertical red line marks the rotation period $P_\mathrm{rot} = 85\,\mathrm{d}$ presented by \citet{Kuzuhara2024}, the horizontal lines represent the 10 per cent (blue) and 1 per cent (orange) false alarm probability levels.}
    \label{fig:longterm-photometry}
\end{figure*}

Fig.~\ref{fig:longterm-photometry} shows the long-term photometry used in Section \ref{sec:starvar} to determine the stellar rotation period.

\section{RV periodograms and data}

\begin{figure*}
    \centering
    \includegraphics[width=\linewidth]{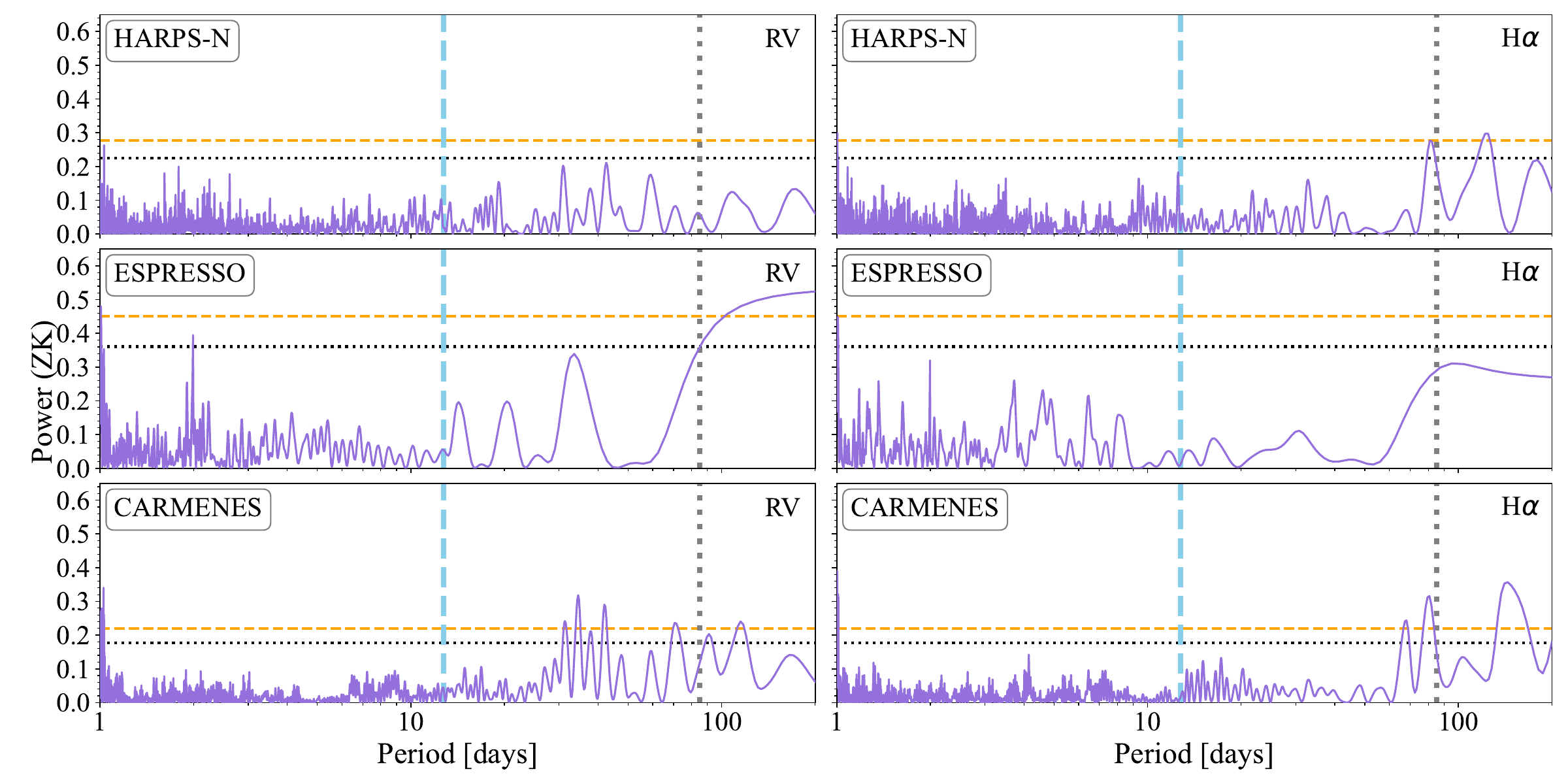}
    \caption{GLS periodograms for the RV and H$\alpha$ data from all three instruments. The vertical lines in the plot are the orbital period of Gliese 12\,b (dashed light blue line, $P = 12.7614\,\mathrm{d}$), and $P_\mathrm{rot} = 85\,\mathrm{d}$ (gray dotted line). The horizontal lines are the false alarm probabilities $(\mathrm{FAP})=10$ per cent (black dotted line) and 1 per cent (dashed orange line).}
    \label{fig:RV_periodograms}
\end{figure*}

Fig.~\ref{fig:RV_periodograms} shows the periodograms for the RV and H$\alpha$ time series for HARPS-N, ESPRESSO, and CARMENES. Table~\ref{tab:rv_data_example} contains a small sample of the RV data used in both the informed RV fit and the joint fit.

\begin{table*}
        \centering
        \caption{A sample of the RV data used in the informed RV fit and the joint fit. The full table is available on Vizier.}
        \renewcommand{\arraystretch}{1.4}
        \begin{tabular}{cccccc}
        \hline
        \hline
            BJD & RV & $\sigma_\mathrm{RV}$ & H$\alpha$ & $\sigma_{\mathrm{H}\alpha}$ & Instrument\\
            (days) & ($\mathrm{m\,s^{-1}}$) & ($\mathrm{m\,s^{-1}}$) & & & \\
            \hline
            2560165.657371 & 51335.99 & 1.42 & -0.00392 & 0.00044 & HARPS-N \\
            2560171.600859 & 51336.51 & 1.69 & 0.02884 & 0.00069 & HARPS-N \\
            2560172.701740 & 51330.45 & 1.39 & -0.00319 & 0.00052 & HARPS-N \\
            2560173.618065 & 51334.65 & 1.26 & 0.01949 & 0.00057 & HARPS-N \\
            2560174.667788 & 51332.18 & 1.28 & 0.00530 & 0.00056 & HARPS-N \\
            ... & ... & ... & ... & ... & ... \\
            2560467.902064 & 51413.63 & 0.30 & -0.00608 & 0.00014 & ESPRESSO \\
            2560473.887225 & 51410.06 & 0.33 & -0.00361 & 0.00017 & ESPRESSO \\
            2560478.903217 & 51409.70 & 0.46 & -0.01765 & 0.00026 & ESPRESSO \\
            2560479.905718 & 51408.93 & 0.29 & -0.00962 & 0.00015 & ESPRESSO \\
            2560486.883796 & 51409.36 & 0.50 & 0.02229 & 0.00030 & ESPRESSO \\
            ... & ... & ... & ... & ... & ... \\
            2560123.593106 & -0.27 & 2.01 & 0.00678 & 0.00331 & CARMENES \\
            2560124.593296 & 1.74 & 2.83 & 0.01442 & 0.00443 & CARMENES \\
            2560125.643064 & -0.69 & 1.41 & -0.00233 & 0.00222 & CARMENES \\
            2560127.644050 & -0.52 & 1.68 & 0.01458 & 0.00207 & CARMENES \\
            2560133.611007 & -1.22 & 2.61 & 0.00517 & 0.00304 & CARMENES \\
            ... & ... & ... & ... & ... & ... \\
            \hline
        \end{tabular}
        \label{tab:rv_data_example}
\end{table*}

\section{Priors and fitted parameters}

\begin{figure*}
    \centering
    \includegraphics[width=\linewidth]{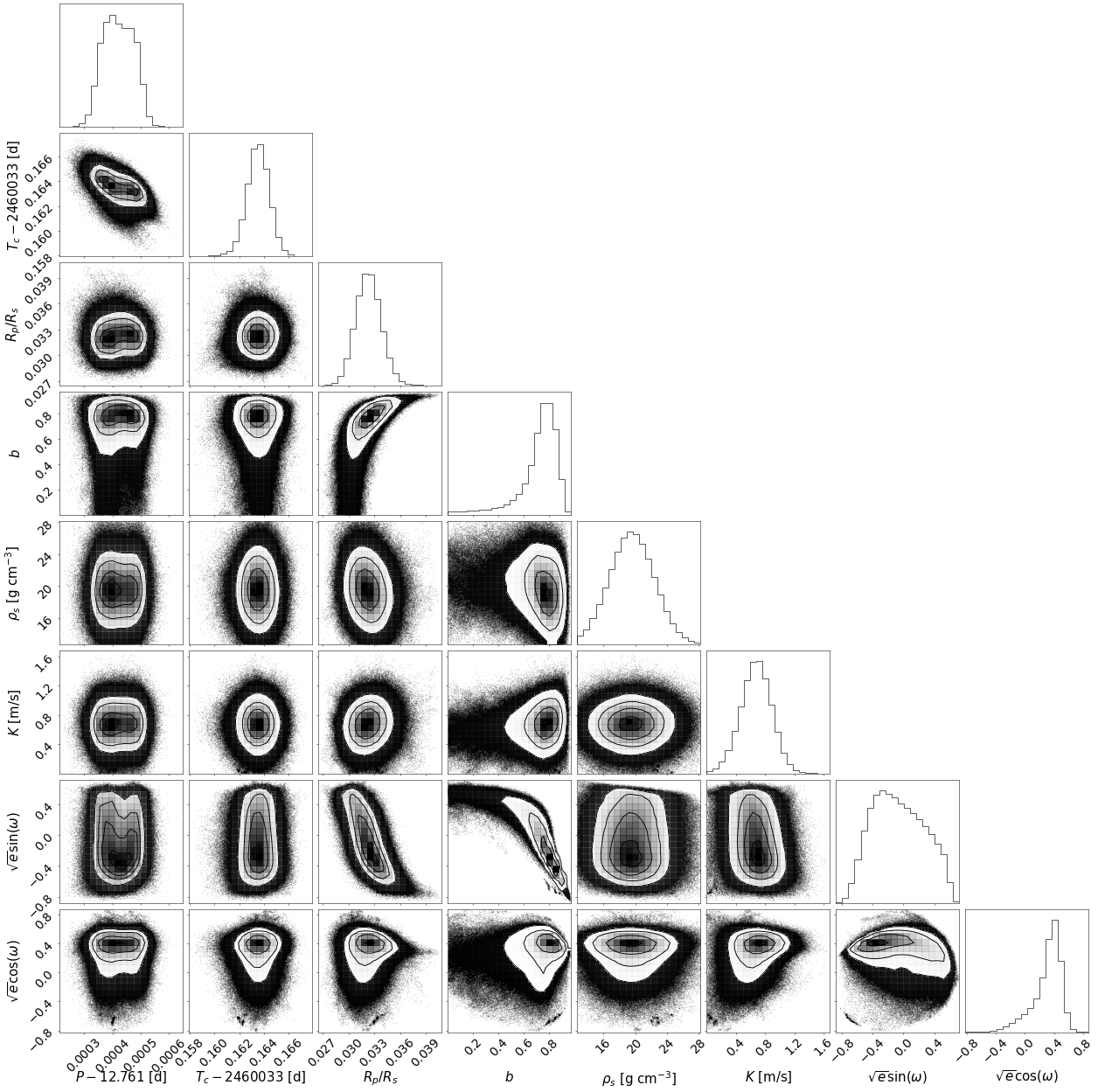}
    \caption{Corner plot for the fitted parameters of the joint fit.}
    \label{fig:corner}
\end{figure*}

Tables \ref{tab:fit_priors_planet} and \ref{tab:fit_priors_starinst} show the prior distributions used during the different fitting procedures. Table \ref{tab:fit_results_starinst} has the final fitted values related to the star as well as the various instruments. The results of the joint fit are presented in Fig.~\ref{fig:corner}, which shows the posteriors of the fitted parameters in a corner plot.

\begin{table*}
        \centering
        \caption{Priors for the planetary parameters.}
        \renewcommand{\arraystretch}{1.4}
        \begin{tabular}{lcccc}
        \hline
        \hline
             Parameter & Unit & Photometry-only fit & Informed RV fit & Joint fit\\
             \hline
             $P$ & (d) & $\mathcal{N}(12.761408,0.1)$ & $\mathcal{N}(12.761422, 0.000047)$ &  $\mathcal{U}(12.46, 13.06)$ \\
             $T_0$ & (BJD) & $\mathcal{N}(2459497.1865,0.1)$ &  $\mathcal{N}(2460492.5747,0.0044)$ & $\mathcal{U}(2460032.86, 2460033.46)$  \\ 
             $R_\mathrm{p}/R_\star$ & --- & $\mathcal{U}(0,1)$ & --- & $\mathcal{U}(0.0,0.5)$ \\
             $b$ & --- & $\mathcal{U}(0,1)$ & --- & $\mathcal{U}(0,1)$ \\
             $\rho_\star$ & (g\,cm$^{-3}$) & $\mathcal{N}(18.9879198,15.950938)$ & --- & $\mathcal{N}(19.317,1.409)$ \\
             $K$ & (m\,s$^{-1}$) & --- & $\mathcal{U}(0,2.5)$ & $\mathcal{U}(0,50)$ \\
             $\sqrt{e}\sin{\omega}$ & --- & --- & $\mathcal{U}(-1,1)$ & $\mathcal{U}(-1,1)$ \\
             $\sqrt{e}\cos{\omega}$ & --- & --- & $\mathcal{U}(-1,1)$ & $\mathcal{U}(-1,1)$ \\
             $e$ & --- & $\mathcal{U}(0,0.9)$  & --- & $\mathcal{U}(0.0,0.95)$ \\
             $\omega$ & (deg) & $\mathcal{U}(0,360)$   & --- &  $\mathcal{U}(0,360)$ \\
             \hline
        \end{tabular}
        \label{tab:fit_priors_planet}
\end{table*}

\begin{table*}
        \centering
        \caption{Stellar and instrumental priors. We note that the coherence GP parameter is different for both fits due to a different functional form of the same kernel.} 
        \renewcommand{\arraystretch}{1.4}
        \begin{tabular}{lcccc}
        \hline
        \hline
             Parameter & Unit & Photometry-only fit & Informed RV fit  & Joint fit\\
             \hline
             $P_\mathrm{rot,GP}$ & (d) & --- & $\mathcal{U}(25,200)$ & $\mathcal{U}(60,200)$ \\
             coherence & --- & --- & $\lambda_\mathrm{p} \in \mathcal{U}(0.01,10)$ & $w \in \mathcal{U}(0.001, 3.0)$ \\
             evolution & (d) & --- & $\lambda_\mathrm{e} \in \mathcal{U}(5, 1000)$ & $\lambda \in \mathcal{U}(120.0, 1000.0)$ \\
             amplitude & (m\,s$^{-1}$) & --- & $A_0 \in \mathcal{U}(0,10)$ & $H, C \in \mathcal{U}(0, 50)$  \\
             jitter HN & (m\,s$^{-1}$) & --- & $\mathcal{U}(0,100)$ & $\mathcal{U}(0,100)$ \\
             jitter CARM & (m\,s$^{-1}$) & --- & $\mathcal{U}(0,100)$ & $\mathcal{U}(0,100)$ \\
             jitter ESPR & (m\,s$^{-1}$) & --- & $\mathcal{U}(0,100)$ & $\mathcal{U}(0,100)$ \\
             offset HN & (m\,s$^{-1}$) & --- & $\mathcal{U}(51230.45, 51446.16)$ & $\mathcal{U}(51000,51800)$ \\
             offset CARM & (m\,s$^{-1}$) & --- & $\mathcal{U}(-106.67, 105.88)$ & $\mathcal{U}(-50,50)$ \\
             offset ESPR & (m\,s$^{-1}$) & --- & $\mathcal{U}(51305.26, 51513.63)$ & $\mathcal{U}(51000,51800)$ \\
             GP$_\sigma$ & (ppm) & $\mathcal{L}(0.000001,1000000)$ & --- & $\log_{10}( \mathcal{U}(-6, 6) ) $ \\
             GP$\rho$ & (d) & $\mathcal{L}(0.001,1000)$ & --- & $\log_{10}( \mathcal{U}(-3, 3) ) $ \\
             LD \textit{TESS} &  & $q_1 \in \mathcal{U}(0,1.0)$ & --- & $u_1 \in \mathcal{N}(0.26,0.07)$ \\
             LD \textit{TESS} & & $q_2 \in \mathcal{U}(0,1.0)$ & --- & $u_2 \in \mathcal{N}(0.29,0.14)$ \\
             LD \textit{CHEOPS} &  & $q_1 \in \mathcal{U}(0,1.0)$ & --- & $u_1 \in \mathcal{N}(0.26,0.04)$ \\
             LD \textit{CHEOPS} & & $q_2 \in \mathcal{U}(0,1.0)$ & --- & $u_2 \in \mathcal{N}(0.28,0.09)$ \\
             LD MuSCAT2 \textit{r} &  & $q_1 \in \mathcal{U}(0,1.0)$ & --- & --- \\
             LD MuSCAT2 \textit{r} & & $q_2 \in \mathcal{U}(0,1.0)$ & --- & --- \\
             LD MuSCAT3 \textit{r} &  & $q_1 \in \mathcal{U}(0,1.0)$ & --- & ---  \\
             LD MuSCAT3 \textit{r} & & $q_2 \in \mathcal{U}(0,1.0)$ & --- & --- \\
             LD MuSCAT2 \textit{i} &  & $q_1 \in \mathcal{U}(0,1.0)$ & --- & --- \\
             LD MuSCAT2 \textit{i} & & $q_2 \in \mathcal{U}(0,1.0)$ & --- & --- \\
             LD MuSCAT3 \textit{i} &  & $q_1 \in \mathcal{U}(0,1.0)$ & --- & --- \\
             LD MuSCAT3 \textit{i} & & $q_2 \in \mathcal{U}(0,1.0)$ & --- & --- \\
             LD MuSCAT2 \textit{z} &  & $q_1 \in \mathcal{U}(0,1.0)$ & --- & --- \\
             LD MuSCAT2 \textit{z} & & $q_2 \in \mathcal{U}(0,1.0)$ & --- & --- \\
             LD MuSCAT3 \textit{z} &  & $q_1 \in \mathcal{U}(0,1.0)$ & --- & --- \\
             LD MuSCAT3 \textit{z} & & $q_2 \in \mathcal{U}(0,1.0)$ & --- & --- \\
             offset \textit{TESS} & & $\mathcal{N}(0,0.1)$ & --- & --- \\
             offset \textit{CHEOPS} &  & $\mathcal{N}(0,0.1)$ & --- & --- \\
             offset MuSCAT2 \textit{r} &  & $\mathcal{N}(0,0.1)$ & --- & --- \\
             offset MuSCAT3 \textit{r} &  & $\mathcal{N}(0,0.1)$ & --- & --- \\
             offset MuSCAT2 \textit{i} &  & $\mathcal{N}(0,0.1)$ & --- & --- \\
             offset MuSCAT3 \textit{i} &  & $\mathcal{N}(0,0.1)$ & --- & --- \\
             offset MuSCAT2 \textit{z} &  & $\mathcal{N}(0,0.1)$ & --- & --- \\
             offset MuSCAT3 \textit{z} &  & $\mathcal{N}(0,0.1)$ & --- & --- \\
             jitter \textit{TESS} & (ppm) & $\mathcal{L}(0.1,1000.0)$ & --- & $\mathcal{U}(0,119800)$; $\mathcal{U}(0,117600)$; $\mathcal{U}(0,118000)$; $\mathcal{U}(0,117600)$;  \\
             & & & --- & $\mathcal{U}(0,128000)$; $\mathcal{U}(0,128500)$; $\mathcal{U}(0,121600)$; $\mathcal{U}(0,121100)$ \\
             jitter \textit{CHEOPS} & (ppm) & $\mathcal{L}(0.1,100000.0)$& --- & $\mathcal{U}(0,69800)$; $\mathcal{U}(0,62700)$; $\mathcal{U}(0,67000)$; $\mathcal{U}(0,63000)$;  \\
             & & & --- & $\mathcal{U}(0,62800)$; $\mathcal{U}(0,251000)$ \\
             jitter MuSCAT2 \textit{r} & (ppm)& $\mathcal{L}(0.1,100000.0)$ & --- & --- \\
             jitter MuSCAT3 \textit{r} & (ppm) & $\mathcal{L}(0.1,100000.0)$ & --- & --- \\
             jitter MuSCAT2 \textit{i} & (ppm) & $\mathcal{L}(0.1,100000.0)$ & --- & --- \\
             jitter MuSCAT3 \textit{i} & (ppm) & $\mathcal{L}(0.1,100000.0)$ & --- & --- \\
             jitter MuSCAT2 \textit{z} & (ppm) & $\mathcal{L}(0.1,100000.0)$ & --- & --- \\
             jitter MuSCAT3 \textit{z} & (ppm) & $\mathcal{L}(0.1,100000.0)$ & --- & --- \\
             \hline
        \end{tabular}
        \label{tab:fit_priors_starinst}
\end{table*}

\begin{table*}
        \centering
        \caption{Stellar and instrumental parameters. We note that the coherence GP parameter is different for both fits due to a different functional form of the same kernel.} 
        \renewcommand{\arraystretch}{1.4}
        \begin{tabular}{lcccc}
        \hline
        \hline
             Parameter & Unit & Photometry-only fit & Informed RV fit & Joint fit\\
             \hline
             $P_\mathrm{rot,GP}$ & (d) & --- & $121.6_{-61.8}^{+51.5}$ & $172.0_{-1.6}^{+1.5}$ \\
             coherence & --- & --- & $\lambda_\mathrm{p} = 3.10_{-2.87}^{+4.64}$ & $w = 0.212_{-0.045}^{+0.054}$  \\
             evolution & (d) & --- & $\lambda_\mathrm{e} = 9.61_{-2.06}^{+56.53}$ & $\lambda = 476_{-160}^{+267}$  \\
             amplitude & (m\,s$^{-1}$) & --- & $A_0 = 2.04_{-0.34}^{+0.44}$ & $H_\mathrm{HN} = 2.24_{-0.65}^{+0.89}$; $H_\mathrm{CARM} = 2.52_{-0.75}^{+1.20}$; $H_\mathrm{ESPR} = 0.84_{-0.15}^{+0.18}$ \\
             & & --- & --- & $C_\mathrm{HN} = 0.78_{-0.55}^{+1.10}$; $C_\mathrm{CARM} = 1.07_{-0.75}^{+1.60}$; $C_\mathrm{ESPR} = 2.2_{-1.6}^{+4.9}$ \\
             jitter HN & (m\,s$^{-1}$) & --- & $1.37_{-0.31}^{+0.32}$ & $1.70_{-0.31}^{+0.34}$ \\
             jitter CARM & (m\,s$^{-1}$) & --- & $0.25_{-0.18}^{+0.29}$ & $0.35_{-0.24}^{+0.35}$ \\
             jitter ESPR & (m\,s$^{-1}$) & --- & $0.84_{-0.14}^{+0.18}$ & $0.84_{-0.15}^{+0.18}$ \\
             offset HN & (m\,s$^{-1}$) & --- & $51337.25 \pm 0.56$ & $51337.18_{-0.92}^{+0.96}$ \\
             offset CARM & (m\,s$^{-1}$) & --- & $0.47_{-0.54}^{+0.56}$ & $0.10_{-1.00}^{+1.20}$ \\
             offset ESPR & (m\,s$^{-1}$) & --- & $51409.56_{-0.58}^{+0.59}$ & $51409.80_{-1.30}^{+1.60}$ \\
             GP$_\sigma$ & (ppm) & $0.000192_{-0.000013}^{+0.000015}$ & --- &  $ \sigma_\mathrm{TESS} = 2.2^{+5.2}_{-1.0}$ ; $ \sigma_\mathrm{TESS} = 2.1^{+4.9}_{-0.9}$ ; $ \sigma_\mathrm{TESS} = 2.3^{+6.4}_{-1.1}$ ; $ \sigma_\mathrm{TESS} = 2.0^{+3.8}_{-0.8}$  \\
              & & --- & --- & $\sigma_\mathrm{TESS} = 2.3^{+5.7}_{-1.1}$ ; $ \sigma_\mathrm{TESS} = 2.2^{+5.5}_{-1.0}$ ; $ \sigma_\mathrm{TESS} = 2.0^{+4.0}_{-0.8}$ ; $ \sigma_\mathrm{TESS} = 2.0^{+4.1}_{-0.8}$ \\
              & & --- & --- & $ \sigma_\mathrm{CHEOPS} = 3.0^{+11.2}_{-1.6}$ ; $ \sigma_\mathrm{CHEOPS} = 3.3^{+20.1}_{-2.0}$ ; $ \sigma_\mathrm{CHEOPS} = 2.6^{+8.1}_{-1.3}$  \\  
              & & --- & --- & $ \sigma_\mathrm{CHEOPS} = 3.2^{+18.7}_{-1.8}$ ; $ \sigma_\mathrm{CHEOPS} = 2.4^{+6.7}_{-1.1}$ ; $ \sigma_\mathrm{CHEOPS} = 2.9^{+11.9}_{-1.6}$ \\ 
             GP$\rho$ & (d) & $0.621_{-0.071}^{+0.075}$ & --- & --- \\
             & & --- & --- & $ \rho_\mathrm{TESS} = 257.0^{+419.0}_{-205.8}$ ; $ \rho_\mathrm{TESS} = 257.0^{+434.8}_{-202.1}$ ; $ \rho_\mathrm{TESS} = 158.5^{+444.1}_{-131.6}$  \\
             & & --- & --- & $ \rho_\mathrm{TESS} = 331.1^{+410.2}_{-249.8}$ $ \rho_\mathrm{TESS} = 208.9^{+436.7}_{-175.8}$ ; $ \rho_\mathrm{TESS} = 204.2^{+441.5}_{-170.3}$ \\
             & & --- & --- & $ \rho_\mathrm{TESS} = 346.7^{+411.8}_{-257.6}$ ; $ \rho_\mathrm{TESS} = 338.8^{+402.5}_{-251.7}$ \\
             & & --- & --- & $ \rho_\mathrm{CHEOPS} = 39.8^{+276.4}_{-35.0}$ ; $ \rho_\mathrm{CHEOPS} = 4.1^{+14.1}_{-2.8}$ ; $ \rho_\mathrm{CHEOPS} = 131.8^{+417.7}_{-116.3}$ \\
             & & --- & --- &  $ \rho_\mathrm{CHEOPS} = 0.2^{+0.7}_{-0.1}$ ; $ \rho_\mathrm{CHEOPS} = 154.9^{+447.7}_{-137.1}$ ; $ \rho_\mathrm{CHEOPS} = 61.7^{+364.9}_{-57.8}$ \\
             LD \textit{TESS} &  & $q_1 = 0.39_{-0.23}^{+0.30}$ & --- & $u_1 = 0.26_{-0.07}^{+0.07}$ \\
             LD \textit{TESS} & & $q_2 = 0.29_{-0.18}^{+0.22}$  & --- & $u_2 = 0.30_{-0.13}^{+0.13}$ \\
             LD \textit{CHEOPS} &  & $q_1 = 0.18_{-0.13}^{+0.19}$ & --- & $u_1 = 0.26_{-0.04}^{+0.04}$ \\
             LD \textit{CHEOPS} & & $q_2 = 0.37_{-0.22}^{+0.28}$ & --- & $u_2 = 0.28_{-0.09}^{+0.09}$ \\
             LD MuSCAT2 \textit{r} &  & $q_1 = 0.23_{-0.16}^{+0.26}$ & --- & --- \\
             LD MuSCAT2 \textit{r} & & $q_2 = 0.42_{-0.27}^{+0.30}$  & --- & --- \\
             LD MuSCAT3 \textit{r} &  & $q_1 = 0.45_{-0.22}^{+0.25}$ & --- & --- \\
             LD MuSCAT3 \textit{r} & & $q_2 = 0.57_{-0.26}^{+0.25}$ & --- & --- \\
             LD MuSCAT2 \textit{i} &  & $q_1 = 0.43_{-0.24}^{+0.23}$ & --- & --- \\
             LD MuSCAT2 \textit{i} & & $ q_2 = 0.36_{-0.24}^{+0.32}$  & --- & --- \\
             LD MuSCAT3 \textit{i} &  & $q_1 = 0.47_{-0.24}^{+0.25}$ & --- & --- \\
             LD MuSCAT3 \textit{i} & & $q_2 = 0.29_{-0.20}^{+0.28}$ & --- & --- \\
             LD MuSCAT2 \textit{z} &  & $q_1 = 0.23_{-0.16}^{+0.28}$ & --- & --- \\
             LD MuSCAT2 \textit{z} & & $ q_2 = 0.40_{-0.26}^{+0.32}$  & --- & --- \\
             LD MuSCAT3 \textit{z} &  & $q_1 = 0.49_{-0.23}^{+0.23}$ & --- & --- \\
             LD MuSCAT3 \textit{z} & & $q_2 = 0.40_{-0.25}^{+0.29}$ & --- & --- \\
             offset \textit{TESS} & & $0.000003_{-0.000021}^{+0.000021}$ & --- & --- \\
             offset \textit{CHEOPS} & & $0.000163_{-0.000018}^{+0.000017}$ & --- & --- \\
             offset MuSCAT2 \textit{r} & & $0.000419_{-0.000046}^{+0.000050}$ & --- & --- \\
             offset MuSCAT3 \textit{r} & & $0.000017_{-0.000031}^{+0.000031}$ & --- & --- \\
             offset MuSCAT2 \textit{i} & & $0.000022_{-0.000034}^{+0.000035}$ & --- & --- \\
             offset MuSCAT3 \textit{i} & & $0.000032_{-0.000034}^{+0.000032}$ & --- & --- \\
             offset MuSCAT2 \textit{z} & & $0.000050_{-0.000044}^{+0.000043}$ & --- & --- \\
             offset MuSCAT3 \textit{z} & & $0.000012_{-0.000034}^{+0.000032}$ & --- & --- \\
             \hline
        \end{tabular}
        \label{tab:fit_results_starinst}
\end{table*}

\begin{table*}
        \contcaption{Stellar and instrumental parameters. We note that the coherence GP parameter is different for both fits due to a different functional form of the same kernel.}
        \centering
        \renewcommand{\arraystretch}{1.4}
        \begin{tabular}{lcccc}
        \hline
        \hline
        Parameter & Unit & Photometry-only fit & Informed RV fit & Joint fit\\
        \hline
             jitter \textit{TESS} & (ppm) & $1.05_{-0.81}^{+5.20}$ & --- & $114_{-71}^{+100}$; $101_{-63}^{+93}$; $390_{-140}^{+110}$; $134_{-85}^{+110}$;  \\
             & & --- & --- & $160_{-100}^{+130}$; $105_{-65}^{+96}$; $99_{-59}^{+90}$; $121_{-76}^{+110}$ \\
             jitter \textit{CHEOPS} & (ppm) & $0.621_{-0.071}^{+0.075}$ & --- & $380_{-52}^{+49}$; $434_{-49}^{+47}$; $356_{-84}^{+75}$; $670_{-66}^{+68}$;  \\
             & & --- & --- & $204_{-110}^{+96}$; $160_{-100}^{+160}$ \\
             jitter MuSCAT2 \textit{r} & (ppm) & $784.91_{-49.42}^{+51.39}$ & --- & --- \\
             jitter MuSCAT3 \textit{r} & (ppm) & $599.73_{-51.95}^{+46.93}$ & --- & --- \\
             jitter MuSCAT2 \textit{i} & (ppm) & $1216.67_{-39.34}^{+40.25}$ & --- & --- \\
             jitter MuSCAT3 \textit{i} & (ppm) & $552.04_{-84.80}^{+71.16}$ & --- & --- \\
             jitter MuSCAT2 \textit{z} & (ppm) & $1335.90_{-44.54}^{+48.65}$ & --- & ---\\
             jitter MuSCAT3 \textit{z} & (ppm) & $405.06_{-385.93}^{+156.39}$ & --- & --- \\
             \hline
        \end{tabular}
        \label{tab:fit_results_starinst_cont}
\end{table*}

\bsp	
\label{lastpage}
\end{document}